\documentclass[twocolumn]{aastex631}

\usepackage[caption=false]{subfig}
\usepackage{threeparttable}

\usepackage{float,graphicx,amsmath,tabularx,booktabs,natbib,threeparttable}

\newcommand{\mypm}{\mathrel{\scalebox{0.8}{$\pm$}}}

\definecolor{dark-red}{rgb}{0.4,0.15,0.15}
\definecolor{dark-blue}{rgb}{0.15,0.15,0.4}
\definecolor{medium-blue}{rgb}{0,0,0.5}
\hypersetup{
    colorlinks, linkcolor={dark-red},
    citecolor={dark-blue}, urlcolor={medium-blue}
}

\newcommand{\beqa}{\begin{eqnarray}} 
\newcommand{\eeqa}{\end{eqnarray}}

\newcommand{\bsub}{\begin{subequations}}
\newcommand{\esub}{\end{subequations}}
\newcommand{\beal}{\begin{align}}
\newcommand{\ealn}{\end{align}}

\newcommand{\kms}{${\rm km~s^{-1}}$}

\newcommand{\Msun}{{\ensuremath{\mathrm{M}_{\odot}}}}

\def \caltech {{Division of Physics, Mathematics and Astronomy, 
California Institute of Technology, Pasadena, CA 91125, USA}}

\def \coo {{Caltech Optical Observatories, California Institute of Technology, Pasadena, CA 91125, USA}}

\def \su {{Department of Astronomy, The Oskar Klein Center, Stockholm University, AlbaNova, 10691 Stockholm, Sweden}}

\def \ipac {{IPAC, California Institute of Technology, 1200 E. California
             Blvd, Pasadena, CA 91125, USA}}

\begin{document}

\title{\Large{Twin peaks: SN~2021uvy and SN~2022hgk in the landscape of double-peaked stripped envelope supernovae}}

\correspondingauthor{Yashvi Sharma}
\email{yssharma@astro.caltech.edu}

\author[0000-0003-4531-1745]{Yashvi Sharma}
\affiliation{\caltech}

\author[0000-0003-1546-6615]{Jesper Sollerman}
\affiliation{\su} 

\author[0009-0001-6903-0131]{William Meynardie}
\affiliation{Department of Astronomy, University of Michigan, 1085 S. University Ave., Ann Arbor, MI 48109, USA}

\author[0000-0002-4223-103X]{Christoffer Fremling}
\affiliation{\coo}

\author[0000-0001-8372-997X]{Kaustav~K.~Das}
\affiliation{\caltech}

\author[0009-0001-3390-5151]{Gene~Yun}
\affiliation{Loomis Laboratory of Physics, University of Illinois at Urbana-Champaign, 1110 W Green St Loomis Laboratory, Urbana, IL, 61801, USA}

\author[0000-0001-5390-8563]{Shrinivas R. Kulkarni}
\affiliation{\caltech} 

\author[0000-0001-6797-1889]{Steve Schulze}
\affiliation{Center for Interdisciplinary Exploration and Research in Astrophysics (CIERA), Northwestern University, 1800 Sherman Ave., Evanston, IL 60201, USA} 

\author[0000-0003-0733-2916]{Jacob Wise}
\affiliation{Astrophysics Research Institute, Liverpool John Moores University, 146 Brownlow Hill, Liverpool L3 5RF, UK}

\author[0000-0003-1325-6235]{Se{\'a}n. J. Brennan}
\affiliation{\su}

\author[0000-0001-5955-2502]{Thomas~G.~Brink}
\affiliation{Department of Astronomy, University of California, Berkeley, CA 94720-3411, USA}

\author[0000-0002-8262-2924]{Michael W. Coughlin}
\affiliation{School of Physics and Astronomy, University of Minnesota, Minneapolis, MN 55455, USA}

\author[0000-0002-5884-7867]{Richard Dekany}
\affiliation{\coo}

\author[0000-0002-3168-0139]{Matthew J. Graham}
\affiliation{\caltech}

\author[0000-0002-0129-806X]{K. R. Hinds}
\affiliation{Astrophysics Research Institute, Liverpool John Moores University, 146 Brownlow Hill, Liverpool L3 5RF, UK}

\author[0000-0003-2758-159X]{Viraj Karambelkar}
\affiliation{\caltech} 

\author[0000-0002-5619-4938]{Mansi M. Kasliwal}
\affiliation{\caltech}

\author[0009-0001-6911-9144]{Maggie L.~Li}
\affiliation{\caltech}

\author[0009-0002-4724-7118]{Kira Nolan}
\affiliation{\caltech}

\author[0000-0001-8472-1996]{Daniel A. Perley}
\affiliation{Astrophysics Research Institute, Liverpool John Moores University, 146 Brownlow Hill, Liverpool L3 5RF, UK}

\author[0000-0003-1227-3738]{Josiah N. Purdum}
\affiliation{\coo}

\author[0000-0003-4725-4481]{Sam Rose}
\affiliation{\caltech}

\author[0000-0001-7648-4142]{Ben Rusholme}
\affiliation{\ipac}

\author[0000-0001-8208-9755]{Tawny Sit}
\affiliation{Department of Astronomy and Center of Cosmology and AstroParticle Physics, The Ohio State University, Columbus, OH 43210, USA}

\author[0000-0003-0484-3331]{Anastasios Tzanidakis}
\affiliation{Department of Astronomy and the DiRAC Institute, University of Washington, 3910 15th Avenue NE, Seattle, WA 98195, USA}

\author[0000-0002-9998-6732]{Avery Wold}
\affiliation{\ipac}

\author[0000-0003-1710-9339]{Lin Yan}
\affil{\coo}

\author[0000-0001-6747-8509]{Yuhan Yao}
\affiliation{Miller Institute for Basic Research in Science, 468 Donner Lab, Berkeley, CA 94720, USA}
\affiliation{Department of Astronomy, University of California, Berkeley, CA 94720-3411, USA}

\begin{abstract}
In recent years, a class of stripped-envelope supernovae (SESNe) has emerged that show two distinct peaks in their light curves, where the first peak cannot be attributed to shock cooling emission. Such peculiar supernovae are often studied individually, explained by invoking some combination of powering mechanisms. However, they have seldom been discussed in the broader context of double-peaked SESNe. In this paper, we attempt to form a picture of the landscape of double-peaked SESNe and their powering mechanisms by adding two more objects -- SN~2021uvy and SN~2022hgk. SN~2021uvy is a broad and luminous SN~Ib with an unusually long rise of the first peak and constant color evolution with rising photospheric temperature during the second peak. Although its first peak is similar to that of SN 2019stc, their second peaks differ in properties, making it unique among double-peaked objects. SN~2022hgk shows striking photometric similarity to SN~2019cad and spectroscopic similarity to SN~2005bf, both of which have been suggested to be powered by a double-nickel distribution in their ejecta. We analyze their light curves and colors, compare them with a sample of other double-peaked published supernovae for which we have additional data, and analyze the light curve parameters of the sample. We observe a correlation (p-value $\sim0.025$) between the peak absolute magnitudes of the first and second peaks. The sample shows variety in the photometric and spectroscopic properties, and thus no single definitive powering mechanism applies to the whole sample. However, sub-groups of similarity exist that can be explained by mechanisms like the double-nickel distribution, magnetar central engine, interaction, and fallback accretion. We also map out the duration between the peaks ($\Delta t^{21}$) vs the difference between peak absolute magnitudes ($\Delta M^{21}$) as a phase-space that could potentially delineate the most promising powering mechanisms for the double-peaked SESNe.
 
\end{abstract}


\section{Introduction} \label{sec:intro}

The number of observed stripped-envelope supernovae (SESNe) showing two distinct light-curve peaks has been increasing in recent years with the advent of wide-field dynamic all-sky surveys. This emerging class of SESNe does not seem to form a homogeneous group, instead, there might be subgroups of objects that share observational similarities and powering mechanisms. A common subgroup is SESNe that show a fast initial decline ($t_{1/2}\lesssim5$ days) and then develop a second (i.e.\ the main) peak that appears like a normal SESN. Such a rapidly declining first peak is often associated with the shock-cooling phase from the extended envelope of the progenitor \citep{PiroNakar2014ApJ...788..193N, Piro2015ApJ...808L..51P,Piro2021ApJ...909..209P} or very nearby circumstellar material \citep[CSM, e.g.,][]{Jin2021ApJ...910...68J,KhatamiKasen2024ApJ...972..140K}, and is commonly observed in Type~IIb SNe \citep{Morales2015MNRAS.454...95M,Pellegrino2023ApJ...954...35P,Crawford2025arXiv250303735C} but also some Type~Ibc SNe \citep{Taddia2016A&A...592A..89T,Das2024}. SESNe with early shock-cooling peaks also appear to show a strong correlation between the first and second peak absolute magnitudes, likely because both peak luminosities are related to the explosion energy \citep[see fig.\,1 of][]{Das2024}. 

However, the rest of the double-peaked SESNe show large heterogeneity in light curve shapes, luminosities, and spectral properties, sometimes varying between the two peaks of the same supernova. Such objects have often been studied individually and compared to a few similar, previously known SNe, and various combinations of powering mechanisms have been invoked. The commonly used powering mechanisms include (see \S\ref{sec:powering} for specific examples, references and discussion): i) double-nickel distribution, ii) delayed magnetar energy injection, iii) interaction with circumstellar material (CSM), iv) energy injection due to fallback accretion, v) eruptive precursor powering the initial peak, and vi) pulsational pair-instability eruptions. For a few SNe, tell-tale signs of the powering mechanism are present in the observations, such as narrow emission lines in the optical spectra (indicating CSM interaction). However, for many double-peaked SESNe, a number of these scenarios, or combinations thereof, can reasonably fit the light curve data well. More seldom have these supernovae been analyzed as a photometric class, but doing so might reveal pockets of homogeneity in this dispersed group of objects that perhaps correlate to a particular powering mechanism. Several of the invoked powering mechanisms also have their own limitations on the brightness of the peaks they can produce, the duration between the two peaks, or other observables that can help differentiate between the mechanisms (see \S\ref{sec:powering}).

Gathering more observations of such peculiar supernovae can be particularly important given the rarity of the objects themselves and the exceptional nature of some of the proposed models. Collecting a larger sample also improves our ability to group these objects systematically based on light-curve similarity. In this paper, we present an extensive analysis of SN~2021uvy -- a bright, slowly evolving double-peaked SN~Ib, and SN~2022hgk -- a moderate luminosity and duration double-peaked SN~Ib, which we have followed as part of the Zwicky Transient Facility (ZTF) survey and were previously mentioned in the sample study by \citet{Das2024}. We compare these two supernovae with a sample of clearly double-peaked, published SNe~Ibc, mainly from the ZTF archive (see \S\ref{sec:sample}). For this double-peaked SESNe sample, we estimate several light-curve parameters and attempt to infer whether any phase space mapped out by these parameters can be useful for discerning the possible powering mechanisms of these objects.

This paper is organized as follows. In \S\ref{sec:obs}, we present the observations of the two SNe and data processing methods. In \S\ref{sec:analysis}, we compare the light curves, colors, bolometric luminosities, and spectra of the two SNe with other similar double-peaked SESNe from the literature. We define a sample of clearly double-peaked SESNe from the ZTF archive in \S\ref{sec:sample}, analyze their light curve parameters in \S\ref{sec:lcparams}, and discuss the landscape of powering mechanisms for this sample in \S\ref{sec:powering}. Finally, we summarize our results in \S\ref{sec:summary}.

\section{Observations}\label{sec:obs}

\subsection{Discovery}\label{sec:obs-discovery}

\subsubsection{SN 2021uvy}
SN~2021uvy (a.k.a.\ ZTF21abmlldj) is located at J2000.0 coordinates $\alpha = 00^h29^m30.88^s$ and $\delta = +12^\circ06\arcmin21\farcs$01 in a faint
host galaxy (SDSS $r$ band 22.2 mag).
The redshift is determined to be $z = 0.0944$ from one of our intermediate resolution spectra at late times (\S\ref{sec:obs-spectra}). SN 2021uvy was first detected in the ZTF survey \citep{Bellm2019b,graham2019,Dekany20} data on 2021 August 4 (MJD 59403.424) at a host-subtracted magnitude of 20.64 in the ZTF $r$ band and was reported \citep{uvy2021TNS} to the Transient Name Server (TNS\footnote{https://www.wis-tns.org}) by the Bright Transient Survey (BTS; \citealp{Fremling2020,Perley2020,Rehemtulla2024}) team. The SN was caught at an early stage 
and has good photometric coverage before and during the rise. It was initially reported as a superluminous supernova (SLSN) candidate \citep{LunnanTNS2021} and initially classified as a SLSN-I at $z\sim0.255$ \citep{PoidevinTNS2021} based on the top SNID \citep{snid2007} match of a spectrum obtained on 2021 August 13 with SPRAT \citep{SPRAT} on the Liverpool Telescope (LT; \citealp{lt}). It was reclassified as SN~Ibc at $z=0.1$ \citep{RidleyTNS2021} by the ePESSTO team \citep{pessto,epessto} using a higher resolution spectrum obtained also on 2021 August 13 with EFOSC2 \citep{efosc2} on ESO's New Technology Telescope, which removed its superluminous candidacy. Finally, it was classified as a SN~Ib-pec (peculiar) by the BTS team \citep{ChuTNS2021} based on a spectrum obtained with the LRIS \citep{Oke1995,Perley2019} spectrograph on the Keck-I telescope on 2021 September 9. SN~2021uvy was interesting as a luminous Type~Ib supernova with an unusually long $\sim50$ days rise to the peak ($M^1_{pk}\approx-19.8$). It became even more peculiar when it brightened again after declining for $\sim25$ days post-peak. The rise of the second peak was also slow ($\sim30$ days) and attained a similar luminosity as the first peak ($M^2_{pk}\approx-19.3$). We obtained follow-up optical imaging and spectroscopic observations until the SN faded below apparent magnitude $m_r=22.7$.

\subsubsection{SN 2022hgk}
SN~2022hgk (a.k.a.\ ZTF22aaezyos) is located at J2000.0 coordinates $\alpha = 14^h10^m23.70^s$ and $\delta = +44^\circ14\arcmin01\farcs$21 in the host galaxy SDSS~J141023.70+441401.8. The redshift is determined to be $z=0.0335$ from a host-galaxy spectrum obtained after the SN faded. SN~2022hgk was first detected in ZTF data on 2022 April 6 (MJD 59675.344) at a host-subtracted 
$r$-band magnitude of 20.76 and reported to TNS \citep{yos2022TNS}. The transient remained fainter than 19 magnitude for the next $\sim25$ days and, thus, was not assigned for follow-up under the BTS survey criteria. Spectroscopic follow-up was triggered only once the transient started brightening again and developed a second peak, and SN~2022hgk was subsequently classified as a SN~Ib by the BTS team \citep{yos2022TNS_classification} based on a spectrum obtained with the SEDM spectrograph \citep{Ben-Ami12,sedm2018} on the Palomar 60-inch telescope on 2022 May 20. We continued follow-up optical imaging and spectroscopy until the SN faded below 21 magnitude.

\subsection{Optical photometry}\label{sec:obs-photometry}

\begin{figure*}[htbp!]
    \centering
    \includegraphics[width=0.9\textwidth]{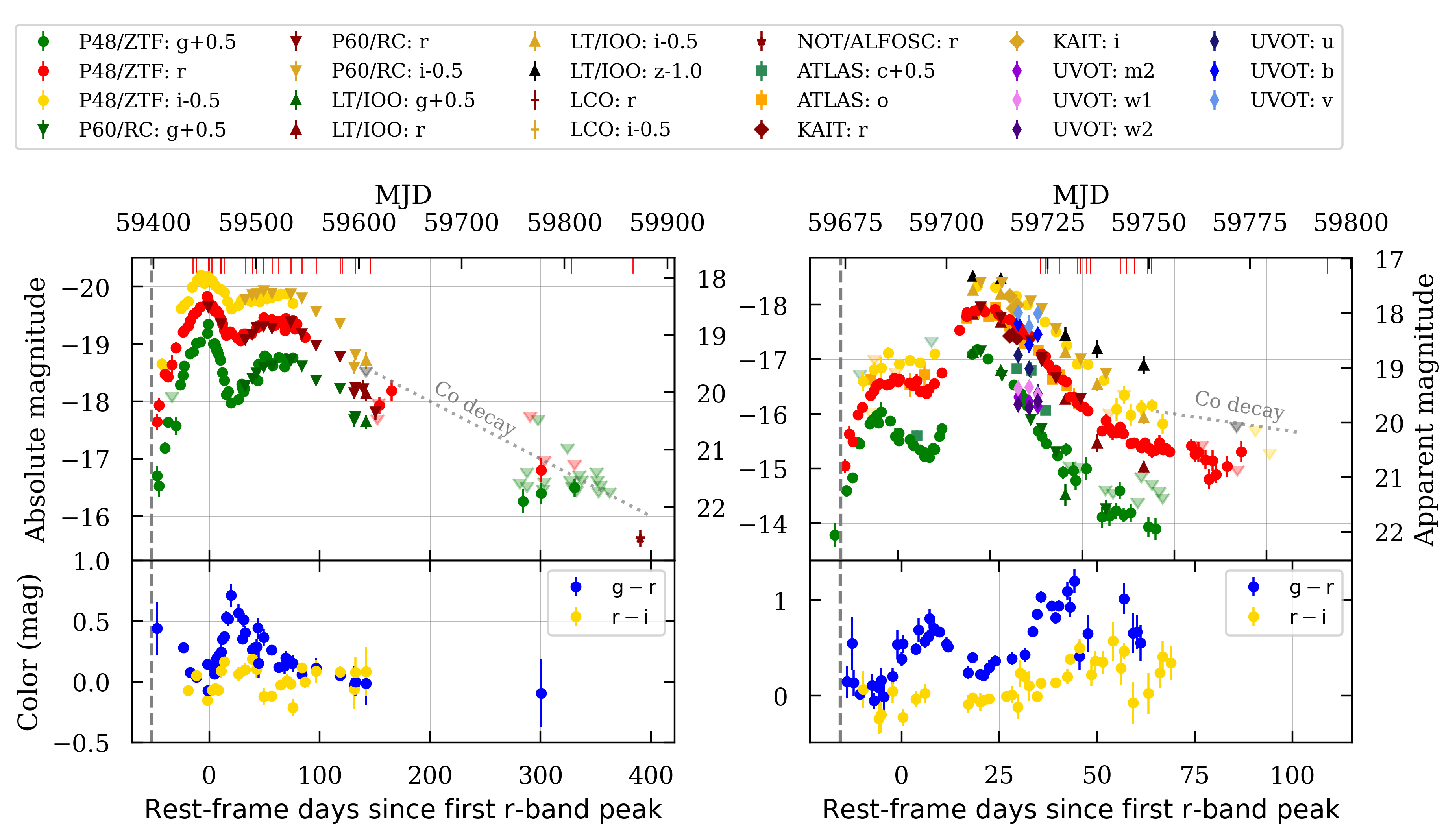}
    \caption{Light (top) and color (bottom) curves of SN~2021uvy (left) and SN~2022hgk (right). The $5\sigma$ detections are shown with solid markers and $3\sigma$ upper limits with transparent markers. All photometry is corrected for MW extinction. Absolute magnitudes are K-corrected (by adding $-2.5log_{10}(1+z)$) and obtained using \citet{Planck2020} cosmology. The $^{56}$Co decay rate (radioactive power) is shown with a dotted gray line. The spectral phases are marked on the top axis with red vertical lines. The explosion epochs are shown with gray dashed lines.}
    \label{fig:lc}
\end{figure*}

For both of these SNe, we obtained forced point-spread function photometry from the ZTF forced photometry service \citep{Masci2019, Masci2023} in $g$, $r$, and $i$ bands and from the ATLAS forced photometry service \citep{atlas, atlassmith2020} in $c$ and $o$ bands. Additional optical photometry was obtained with the Rainbow camera on the Palomar 60-inch telescope \citep{Cenko2006}, the Optical wide-field camera (IO:O) on LT, ALFOSC on the Nordic Optical Telescope (NOT), and the imaging camera on the Katzman Automatic Imaging Telescope (KAIT) at Lick Observatory. The data from P60 and KAIT were processed with the automatic image subtraction pipeline FPipe \citep{Fremling2016}. The data from LT were processed with custom image subtraction and analysis software (K. Hinds and K. Taggart et al., in prep.), and the photometry was measured using PSF fitting techniques from \citet{Fremling2016}. The data from NOT were reduced with PyNOT\footnote{\href{https://github.com/jkrogager/PyNOT}{https://github.com/jkrogager/PyNOT}} data reduction pipeline, image subtraction to remove host contribution was performed with \texttt{HOTPANTS} version 5.11 \citep{Becker2015a} using a pre-supernova $r$ band image from the DESI Legacy Imaging Surveys \citep[LS;][]{Dey2019a}, and the aperture photometry was calibrated against a set of stars from the DESI Legacy Imaging Surveys. 

All photometry presented in this paper is corrected for Milky Way extinction using the Python package extinction \citep{extinction2016}, the dust extinction law from \citet{fitzp1999}, the \citet{Schlafly_2011} dust map, $E(B-V) = 0.067$ mag for SN~2021uvy, $E(B-V) = 0.005$ mag for SN~2022hgk, and $R_V = 3.1$ for both SNe. All measurements are converted into flux units for the analysis. The luminosity distances (and in turn, distance moduli and absolute magnitudes) are calculated using the cosmology parameters from \citet{Planck2020} ($H_0=67.7$, $\Omega_m=0.31$, $\Omega=1$). The absolute magnitudes are calculated using a distance modulus (DM) of 38.254 for SN~2021uvy and 35.879 for SN~2022hgk and are K-corrected. Given the absence of \ion{Na}{1}D narrow absorption in spectra of both SNe and the faint host galaxy of SN~2021uvy, we do not account for any host reddening. The optical photometry data are included in Appendix~\ref{app:phot} and shown in Figure~\ref{fig:lc}.

\subsection{\textit{Swift} Ultraviolet/Optical telescope photometry}

The field of SN~2022hgk was observed with the Ultraviolet/Optical Telescope \citep{Roming2005a} (UVOT) aboard the \textit{Swift} satellite \citep{Gehrels2004a} between MJD $=59720.72$ and 59732.38 in bands $w2$, $m2$, $w1$, $u$, $b$, and $v$. We retrieved science-ready data from the \textit{Swift} archive\footnote{\url{https://www.swift.ac.uk/swift_portal}}. The all-sky exposures for a given epoch and filter were co-added to boost the signal-to-noise ratio using \texttt{uvotimsum} in HEAsoft\footnote{\url{https://heasarc.gsfc.nasa.gov/docs/software/heasoft/}} version 6.31.1. We measured the brightness of the SN with the \textit{Swift} tool {\tt uvotsource}, setting the source aperture radius of $5''$ and a significantly larger background region. All measurements were calibrated with the latest calibration files and converted to the AB system following \citet{Breeveld2011a}. The UV photometry (not host-subtracted) is included in Appendix~\ref{app:phot}. Since the UV photometry is not corrected for host contribution, we do not use it to construct the bolometric light curves.

\subsection{Optical spectroscopy}\label{sec:obs-spectra}
\begin{figure*}[htbp!]
    \centering
    \includegraphics[width=0.6\textwidth]{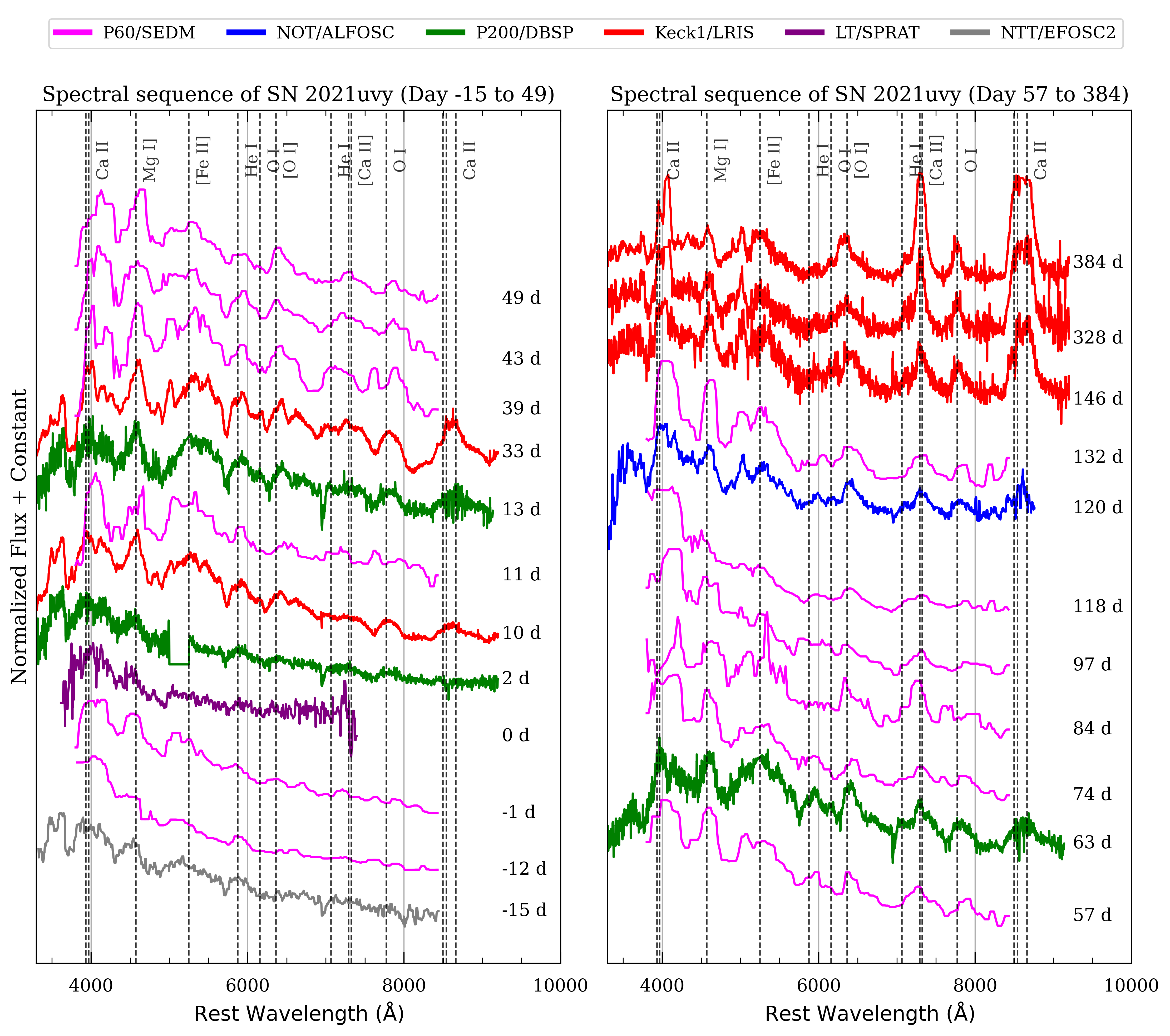}
    \includegraphics[width=0.315\textwidth]{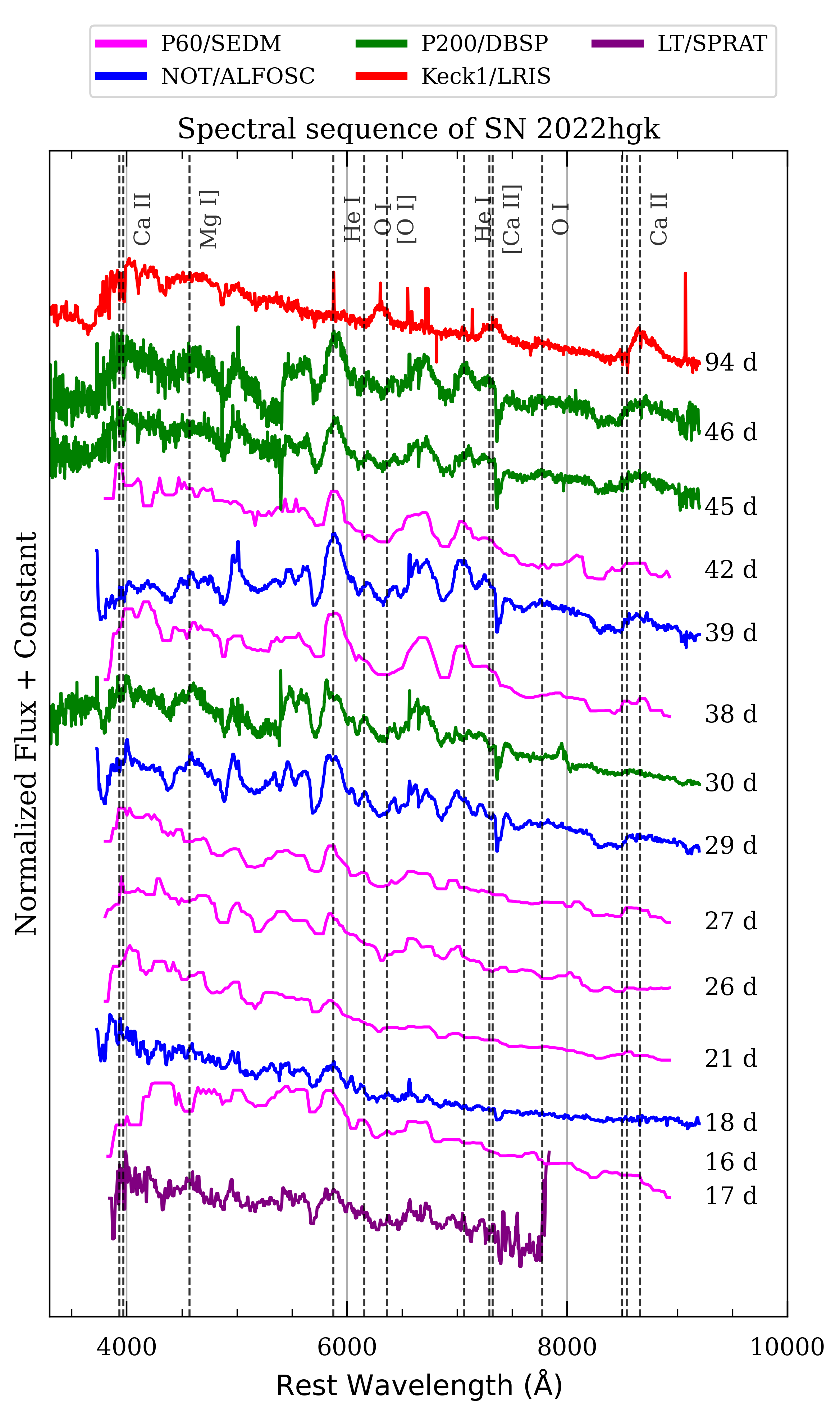}
    \caption{Spectral sequences of SN~2021uvy (left and center) covering epochs from $-15$ to $384$ rest-frame days since its first peak and of SN~2022hgk (right) covering epochs from 17 to 95 rest-frame days since its first peak. Some characteristic spectral lines are marked with vertical gray dashed lines. Spectra are smoothed with a median filter of window size 5.}
    \label{fig:specseq}
\end{figure*}

We obtained spectroscopic follow-up for SN~2021uvy between 2021 August 16 and 2023 July 23 and for SN~2022hgk between 2022 May 8 and 2022 July 27 with the following instruments:
\begin{itemize}
    \item Spectral Energy Distribution Machine (SEDM, $R \sim 100$, \citealp{sedm2018}) on P60, data processed using \texttt{pysedm} \citep{pysedm2019, Kim2022}
    \item Low Resolution Imaging Spectrometer (LRIS, $R \sim 800–1400$, \citealp{Oke1995}) on the Keck-I telescope, data processed using \texttt{LPipe} \citep{Perley2019}
    \item Double Beam Spectrograph (DBSP, $R \sim 1000$, \citealp{Oke1982}) on the Palomar 200-inch telescope (P200), data processed using \texttt{DBSP-DRP} \citep{Roberson2022, pypeit:joss_pub}
    \item Alhambra Faint Object Spectrograph and Camera (ALFOSC, $R \sim 360$), on the Nordic Optical Telescope (NOT), data processed using \texttt{PypeIt} \citep{pypeit:joss_pub}
    \item Spectrograph for the Rapid Acquisition of Transients (SPRAT, $R \sim 360$, \citealp{SPRAT}) on the Liverpool Telescope (LT, \citealp{lt}). Data processed using a custom Python pipeline utilizing the packages Astropy \citep{astropy2022}, NumPy \citep{numpy}, SciPy \citep{scipy2020}, and Matplotlib \citep{matplotlib}.
\end{itemize}  

We present 23 spectra of SN~2021uvy in this paper (22 from the ZTF group, 1 from TNS,  \citealp{RidleyTNS2021}) covering epochs from $-15$ to 384 rest-frame days from its first peak in the $r$ band and 14 spectra of SN~2022hgk covering epochs from 17 to 95 rest-frame days from its first peak in the $r$ band. The spectral sequences are listed in Appendix~\ref{app:spec} and shown in Figure~\ref{fig:specseq}. We also present spectra obtained as part of the ZTF follow-up campaigns of double-peaked SESNe in our sample (see \S\ref{sec:sample} for details) that have not been published previously in Appendix~\ref{app:spec}, namely SN~2020acct (11 spectra, $-1$ to 149 rest-frame days), SN~2021pkd (4 spectra, $-7$ to 7 days), and SN~2023plg (22 spectra, 70 to 147 days). All spectra were corrected for Milky Way extinction using the same procedure as the photometry, then calibrated using contemporaneous host-subtracted ZTF data in the $r$ band. All spectra will be made available on WISeREP \citep{wiserep}. 
\section{Analysis}\label{sec:analysis}

\subsection{The double-peaked SESN sample}\label{sec:sample}

\begin{table*}[htbp!]
    \centering
    \caption{Sample of published double-peaked SESNe in the ZTF archive}
    \label{tab:sample}
    \begin{threeparttable}
        \begin{tabularx}{0.95\textwidth}{ccccccc}
            \toprule
            \toprule
            IAU Name & ZTF Name  & Redshift & Type & $E(B-V)_{MW}$ & $E(B-V)_{host}$ & Reference \\
             &  & & & (mag) & (mag) & \\
             \midrule
    SN~2018ijp  & ZTF18aceqrrs & 0.0848  & Ic    & 0.03 & 0.0           & \citet{Tartaglia2021} \\ 
    SN~2019cad  & ZTF19aamsetj & 0.02751 & Ic    & 0.02 & 0.5\tnote{a}  & \citet{Gutierrez2021} \\
    SN~2019oys  & ZTF19abucwzt & 0.0162  & Ib    & 0.09 & 0.0           & \citet{Sollerman2020}\\
    SN~2019stc  & ZTF19acbonaa & 0.1178  & Ic/SLSN-I    & 0.08 & 0.18\tnote{b} & \citet{Gomez2021} \\
    SN~2020acct & ZTF20acwobku & 0.0347  & Ibc   & 0.03 & 0.0           & \citet{Angus2024} \\
    SN~2021pkd  & ZTF21abfjlxb & 0.0398  & Ib    & 0.04 & 0.0           & \citet{Soraisam2022} \\
    SN~2021uvy  & ZTF21abmlldj & 0.0944  & Ib    & 0.07 & 0.0           & \citet{Das2024} \\
    SN~2022hgk  & ZTF22aaezyos & 0.0335  & Ib    & 0.01 & 0.0           & \citet{Das2024} \\
    SN~2022jli  & ZTF22aapubuy & 0.0055  & Ic    & 0.04 & 0.25\tnote{c} & \citet{Chen2024Natur.625..253C} \\
    SN~2022xxf  & ZTF22abnvurz & 0.0034  & Ic-BL & 0.04 & 0.8\tnote{d}  & \citet{Kuncarayakti2023} \\
    SN~2023aew  & ZTF23aaawbsc & 0.025   & Ibc   & 0.04 & 0.0           & \citet{Kangas2024}\tnote{e} \\
    SN~2023plg  & ZTF23aaxuvkn & 0.027   & Ibc   & 0.06 & 0.0           & \citet{Sharma2024} \\
             \bottomrule
        \end{tabularx}
        \begin{tablenotes}
        \item[a] From \citet{Gutierrez2021}
        \item[b] From \citet{ChenYan2023}
        \item[c] From \citet{Chen2024Natur.625..253C}
        \item[d] From \citet{Kuncarayakti2023}       
        \item[e] Also \citet{Sharma2024}
        \end{tablenotes}
        \hrule
    \end{threeparttable}
\end{table*}

To collect the sample of previously published double-peaked SESNe in ZTF, we looked at ZTF light curves of all unambiguously classified SESNe (Type~Ib, Ic, Ic-BL, Ib/c, Ib-pec) in the ZTF archive (a total of 501 objects). We obtained the light curves from \texttt{Fritz} \citep{skyportal2019,Coughlin2023} and interpolated them using Gaussian process regression. We then used \texttt{scipy.signal.find\_peaks} functionality to search for prominent peaks in the $r$-band light curves (and $g$-band light curves in cases where $r$-band data were not available). We visually vetted the light curves that were identified to have $>1$ peak (46 out of 501) and rejected objects that: i) had incorrect identification of multiple peaks due to missing coverage (and consequently incorrect interpolation), ii) had more than two bumps/peaks (for example, SN~2021efd identified as a bumpy SN in \citealt{Soraisam2022}), or iii) had non-prominent bumps and plateaus. We also rejected one object that fits the double-peak criteria but has not yet been published. SN~2022jli did not get filtered out with this methodology, as its first peak was not covered in ZTF, but we added it to our sample since it is a known peculiar double-peaked supernova. The resulting sample (12 SNe) is summarized in Table~\ref{tab:sample}, and includes SNe~2021uvy and 2022hgk. We note that SN~2019stc is classified as a luminous SN~Ic in \citet{Gomez2021}, but if host extinction is considered, it reaches superluminous status and is classified as a SLSN-I in \citet{ChenYan2023}. Thus, the observed fraction of clearly double-peaked Type Ibc SNe is $\sim2.5\%$ of all Type~Ibc SNe. In the following sections, we compare the photometric and spectroscopic properties of our two key objects, SNe~2021uvy and 2022hgk, with supernovae from this collected sample.

\subsection{Light curves}

We fit the rise of SN~2021uvy in ZTF data with an exponential curve to constrain the explosion epoch, as the rise time is unusually long, but for SN~2022hgk, we fit the rise with a power-law curve. We converted the $r$, $g$, and $i$-band magnitudes into linear flux densities (in $\mu$J), then used the Markov Chain Monte Carlo (MCMC) technique with the following equation to fit the exponential rise in the bands separately:
\begin{equation}
    f = f_{max}(1-e^{\frac{(t_{exp}-t)}{t_c}})
\end{equation}
where $f$ is the flux in $\mu$Jy, $t_{exp}$ is the explosion epoch, $f_{max}$ is the maximum flux, and $t_c$ is the characteristic rise-time. We then calculate the mean and standard deviation of the best-fit values of $t_{exp}$ obtained from the three ZTF bands to get the final explosion epoch at MJD $59398.21 \pm 2.50$ for SN~2021uvy and MJD $59673.80 \pm 4.60$ for SN~2022hgk. 

Figure~\ref{fig:lc} shows the light- (top panel) and color- (bottom panel) curves of SN~2021uvy (left) and SN~2022hgk (right). Both objects show very conspicuous double-peaked light curves, which is highly unusual for SESNe. There are also obvious differences in luminosities and timescales. SN~2021uvy's first peak is broad and has a very slow rise of 52 rest-frame days from explosion to a peak absolute magnitude of $M^1_{pk,r} = -19.8$ in the $r$ band. It then declines for 25 rest-frame days at a rate of $0.030 \pm 0.002$\,mag\,d$^{-1}$ in the $r$ band, faster than the radioactive Co-decay rate ($\approx0.01$\,mag\,d$^{-1}$). After a clear minimum at around MJD 59480, SN~2021uvy brightens again for $\sim28$ rest-frame days to an absolute magnitude of $M^2_{pk,r} = -19.3$ (slightly fainter than the first peak), then slowly declines at a rate of $0.011 \pm 0.001$ \,mag\,d$^{-1}$, very close to the decay rate of $^{56}$Co, shown by the gray dotted line in Figure~\ref{fig:lc} (left). 

On the other hand, SN~2022hgk is nearly two magnitudes fainter at maximum luminosity than SN~2021uvy, has an overall shorter duration and a more luminous second peak compared to the first peak, unlike SN~2021uvy. SN~2022hgk has a first rise time of $\sim16$ rest-frame days from explosion to a peak absolute magnitude of $M^1_{pk,r} = -16.6$, after which it slightly declines for only $\sim7$ rest-frame days before brightening again to a peak absolute magnitude of $M^2_{pk,r} = -17.9$. The peak-to-peak duration ($\Delta t^{21}$, more details in \S\ref{sec:lcparams}) in the $r$ band for SN~2022hgk is $\sim22$ rest-frame days compared to $\sim66$ days for SN~2021uvy. The final decline of SN~2022hgk proceeds at a rate of $0.078 \pm 0.002$\,mag\,d$^{-1}$ in the $r$ band until around MJD 59750, after which the decline appears to become slower and similar to the Co decay rate. 

\begin{figure}[h]
    \centering
    \includegraphics[width=0.48\textwidth]{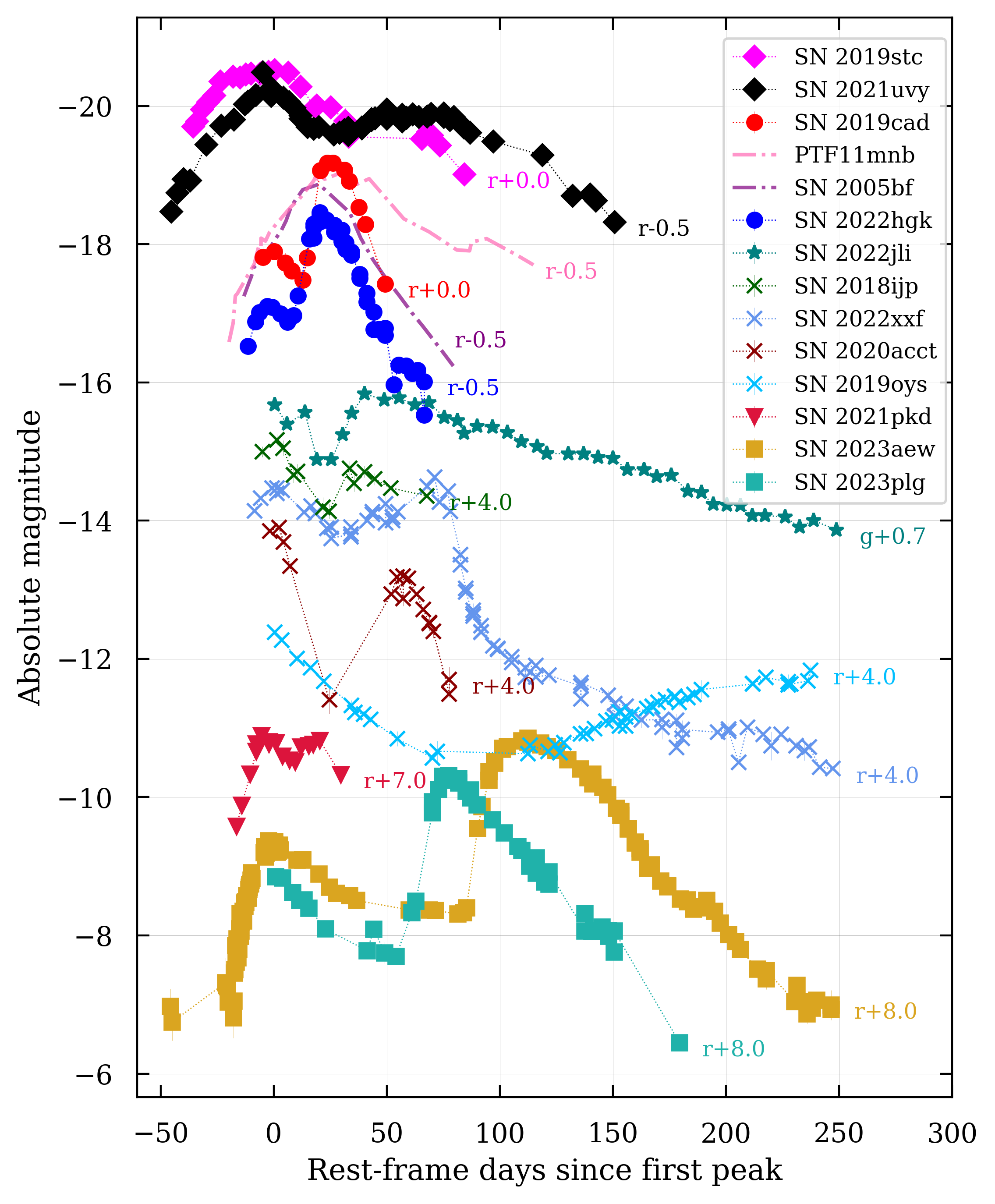}
    \caption{Light curves ($r$-band) of our sample of double-peaked SESNe, shifted vertically for clarity and with their first peaks aligned. Also shown for comparison are SN~2005bf and PTF11mnb (dashdot lines). All absolute magnitudes have been calculated using the same cosmology.}
    \label{fig:lcr}
\end{figure}

In Figure~\ref{fig:lcr}, we show the absolute $r$-band light curves of our double-peaked sample, along with the $r$-band light curves of peculiar double-peaked SNe like SN~2005bf (Type Ib; \citealp{Anupama2005,Folatelli2006,Maeda2007}) and PTF11mnb (Type Ic; \citealp{Taddia2018}). We obtained the light curves of SNe in our sample following \S\ref{sec:obs-photometry} and binned them into 3-day bins. The absolute magnitudes of all SNe shown were calculated using the same cosmology (see \S\ref{sec:obs-photometry}), and host-galaxy extinction was taken into account wherever available. The light curves have been shifted horizontally to align their first peaks and shifted vertically for clarity. 

\begin{table*}[htbp!]
    \centering
    \caption{Light curve parameters of our double-peaked SESN sample. The rise and fade times are calculated between peak flux and half-of-peak flux. The superscripts `1' and `2' denote the first and second peak parameters, respectively. The rise times, fade times, and duration between the two peaks ($\Delta t^{21}$) are reported in rest-frame days.}
    \footnotesize
    \begin{threeparttable}
    \begin{tabularx}{0.99\textwidth}{cccccccccccc}
    \toprule
    \toprule
      SN & Band & $t^1_{rise,1/2}$ & $M^1_{pk}$ & MJD$^1_{pk}$ & $t^1_{fade,1/2}$ & $t^2_{rise,1/2}$ & $M^2_{pk}$ & MJD$^2_{pk}$ & $t^2_{fade,1/2}$ & $\Delta t^{21}$ & $\Delta M^{21}$ \\
       & & (days) & (mag) &  & (days) & (days) & (mag) & & (days) & (days) & (mag) \\
    \midrule
       2018ijp & $r$ & 8.1$\mypm$0.3 & $-$19.16$\mypm$0.13 & 58438 & $>9.2$ & $>18.4$ & $-$18.67$\mypm$0.08 & 58481 & 33.7$\mypm$1.0 & 39.6 & 0.49 \\
       2019cad & $r$ & $>7.8$ & $-$17.87$\mypm$0.09 & 58567 & $>7.8$ & 7.7$\mypm$0.2 & $-$19.17$\mypm$0.03 & 58593 & 13.2$\mypm$0.5 & 25.3 & $-1.30$\\
       2019oys & $r$ & - & $-$16.35$\mypm$0.05 & 58723 & 22.7$\mypm$0.9 & 111.6$\mypm$1.4 & $-$15.74$\mypm$ 0.02 & 58982 & 277.5$\mypm$12.7 & 254.9 & $>0.6$ \\
       2019stc & $r$ & 34.0$\mypm$0.7 & $-$20.52$\mypm$0.05 & 58799 & 29.3$\mypm$1.0 & $>20.6$ & $-$19.60$\mypm$0.06 & 58876 & $>17.0$ & 68.9 & 0.92\\
       2020acct & $r$ & $>2.9$ & $-$18.06$\mypm$0.03 & 59196 & 6.6$\mypm$0.4 & 3.9$\mypm$0.1 & $-$17.21$\mypm$0.01 & 59253 & 13.7$\mypm$0.4 & 55.1 & 0.85 \\
       2021pkd & $r$ & 12.0$\mypm$0.6 & $-$17.84$\mypm$0.06 & 59394 & $>6.7$ & $>10.6$ & $-$17.80$\mypm$0.05 & 59414 & $>12.5$ & 19.2 & 0.04 \\
       2021uvy  & $g$ & 22.6$\mypm$1.0 & $-$19.80$\mypm$0.09 & 59455 & 10.8$\mypm$0.9 & 43.9$\mypm$0.1 & $-$19.24$\mypm$0.02 & 59531 & 56.1$\mypm$0.6 & 69.4 & 0.56\\
         & $r$ & 25.6$\mypm$1.9 & $-$19.77$\mypm$0.08 & 59455 & 20.0$\mypm$0.7 & $>41.1$ & $-$19.37$\mypm$0.01 & 59528 & 55.6$\mypm$0.4 & 66.7 & 0.40 \\
       2022hgk  & $g$ & 5.5$\mypm$0.4 & $-$16.44$\mypm$0.05 & 59684 & 11.9$\mypm$0.4 & 6.4$\mypm$0.4 & $-$17.63$\mypm$0.02 & 59712 & 10.7$\mypm$0.3 & 27.1 & $-1.20$\\
         & $r$ & 11.8$\mypm$0.7 & $-$16.61$\mypm$0.13 & 59691 & $>5.8$ & 7.7$\mypm$0.1 & $-$17.92$\mypm$0.01 & 59713 & 14.8$\mypm$0.5 & 21.3 & $-1.31$\\
       2022jli & $g$ & - & $-$16.37$\mypm$0.01 & 59708 & $>17.9$ & 12.6$\mypm$0.7 & $-$16.54$\mypm$0.04 & 59750 & 73.0$\mypm$3.9 & 41.8 & $>-0.2$ \\
       2022xxf & $r$ & $>8.0$ & $-$18.47$\mypm$0.01 & 59880 & $>24.9$ & 33.9$\mypm$2.1 & $-$18.66$\mypm$0.02 & 59950 & 9.0$\mypm$0.1 & 69.8 & $-$0.19 \\
       2023aew & $r$ & 11.7$\mypm$0.1\tnote{a} & $-$17.28$\mypm$0.01 & 59959 & 34.6$\mypm$0.5 & 19.5$\mypm$0.1 & $-$18.84$\mypm$0.01 & 60075 & 32.2$\mypm$0.1 & 113.2 & $-$1.50\\
       2023plg & $r$ & - & $-$16.83$\mypm$0.02 & 60170 & $>22.1$ & 7.8$\mypm$0.1 & $-$18.30$\mypm$0.02 & 60249 & 23.1$\mypm$0.6 & 76.9 & $>-$1.5 \\
         \bottomrule
    \end{tabularx}
    \begin{tablenotes}
        \item[a] Derived from TESS-Red band data  
        \\
    \end{tablenotes}
        \hrule
    \end{threeparttable}
    \label{tab:lcpars}
\end{table*}

Immediately, we can deduce from Figure~\ref{fig:lcr} that there is significant diversity across the sample, but also sub-groups that share some light curve properties. The slow rise, peak luminosity, and first decline of SN~2021uvy are similar to what is seen for SN~2019stc, a luminous SESN \citep{Gomez2021,ChenYan2023}. \citet{Gomez2022} mentions that the first peaks of both SNe~2019stc and 2021uvy fit well to a combined magnetar central engine and $^{56}$Ni radioactive decay power model, but have weaker magnetar engines than typical SLSNe. They posit that this could explain the SLSNe-like light curve but normal SESNe-like spectra of SN~2019stc. However, this combined model does not account for the rebrightening and cannot explain the second peaks of these two SNe. 
SN~2022hgk's $r$-band light curve and color curve are remarkably similar to those of SN~2019cad, also considered analogous to SN~2005bf and PTF11mnb. The luminosities and timescales of the two peaks of this group of objects, especially the initial rise before the first peak, which is $>10$ days from the explosion, fit the double-nickel distribution scenario \citep{Folatelli2006,Bersten2013,Orellana2022A&A...667A..92O} well. The final declines of these objects have some variation, with PTF11mnb and SN~2022hgk possibly showing a bump toward the end. The group of SESNe with confirmed CSM interaction signatures (SN~2018ijp -- hydrogen-rich dense shell, SN~2019oys -- hydrogen-rich CSM and high-ionization coronal lines, SN~2020acct -- narrow emission lines during first peak, and SN~2022xxf -- late-time narrow emission lines) are shown with crosses in Figure~\ref{fig:lcr} and display the most variety in their light curve evolution, with some having ultra-long durations than others. The accretion-powered SN~2022jli is entirely unique, showing periodic undulations in its long decline. SN~2023aew and 2023plg both have widely separated peaks with a plateau connecting the two peaks and appear unlike any of the other SNe in the sample. Finally, SN~2021pkd does not share a strong similarity with any of the other SNe. 

Table~\ref{tab:lcpars} shows the light-curve parameters (luminosities at both peaks, rise and decline times in different filters measured from peak flux to half of the peak flux) for the double-peaked SESN sample. These parameters were all consistently estimated from interpolated ZTF light curves of the listed SNe when available (TESS-Red band data from \citealt{Sharma2024} were used for the first-peak of SN~2023aew, and ASAS-SN $g$-band data from \citealt{Chen2024Natur.625..253C} were used for the first-peak of SN~2022jli). The interpolation was performed using Gaussian process regression with the help of the \texttt{HAFFET} Python package \citep{YangHaffet}. We are collecting all these parameters in order to map out the landscape of double-peaked SESNe in terms of observable properties, and the ranges and distributions of these properties might later be valuable to constrain the viable powering mechanisms for their light curves. The grouping seen in Figure~\ref{fig:lcr} is also apparent from this table, with some groups (e.g., SNe~2019stc, 2021uvy) having long rest-frame duration between the two peaks ($\Delta t^{21}$) and a fainter second peak ($\Delta M^{21} = M^2_{pk}-M^1_{pk} > 0$), while others (e.g., SNe~2019cad, 2022hgk) having shorter $\Delta t^{21}$ and brighter second peak ($\Delta M^{21} < 0$). SNe~2023aew and 2023plg sit independently in this phase space, with a longer duration like the first group and a brighter second peak like the second. 

\subsection{Bolometric luminosities}

\begin{figure}[h]
    \centering
    \includegraphics[width=0.48\textwidth]{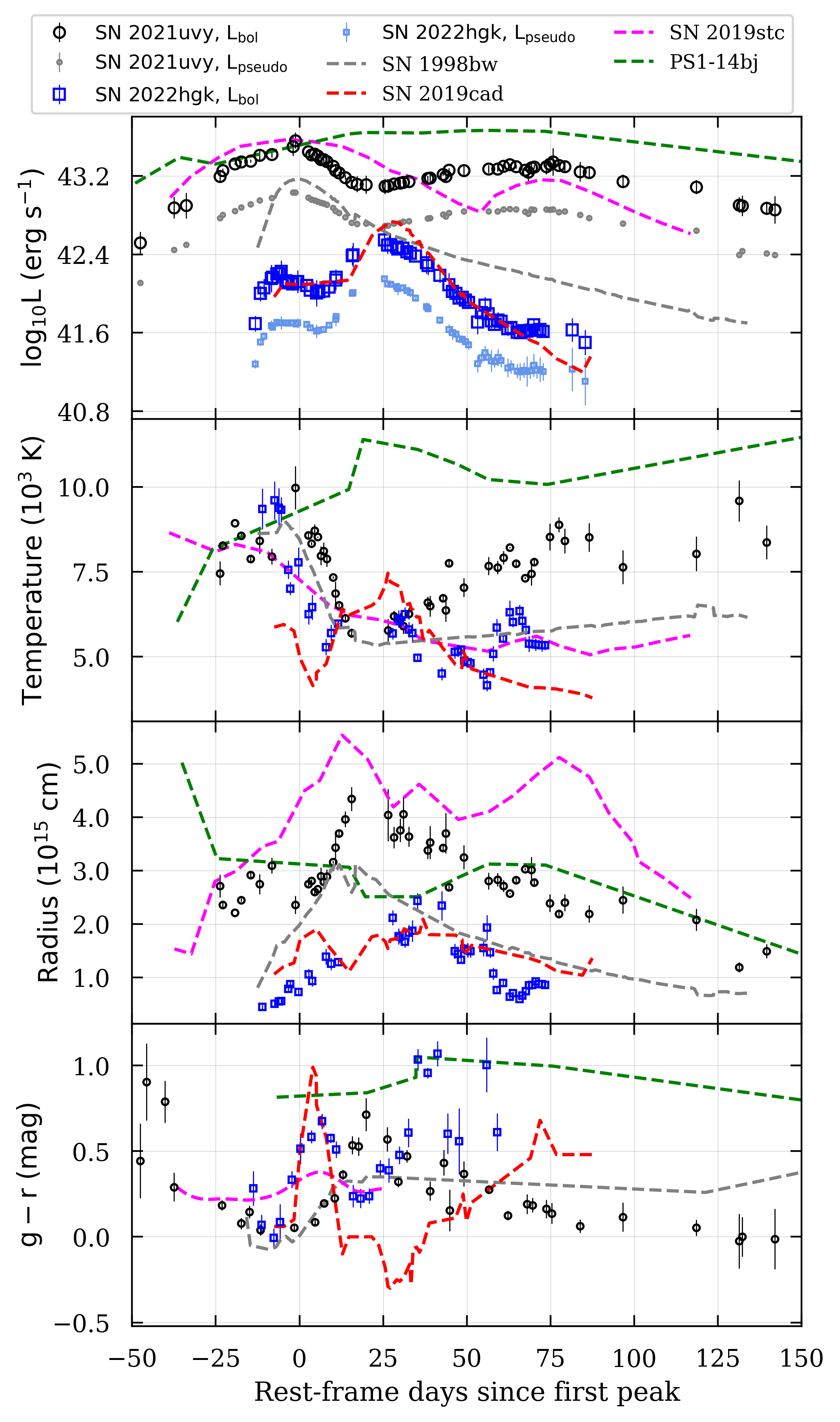}
    \caption{Evolution of the bolometric luminosity (top), blackbody temperature (second), blackbody radius (third), and $g-r$ color (bottom) with time of SN~2021uvy (black circles) and SN~2022hgk (blue squares). Shown for comparison are SN~2019cad (red), SN~2019stc (magenta), PS1-14bj (green), and SN~1998bw (gray). SN~2021uvy and SN~2019stc show similar evolution during their first light-curve peaks but diverge in behavior during the second peaks. SN~2021uvy develops a gradual rise in temperature during the second peak, similar to PS1-14bj, which also correlates with the lack of color evolution for both these SNe. The properties of SN~2022hgk closely resemble those of SN~2019cad.}
    \label{fig:lbol}
\end{figure}

We used \texttt{Superbol} \citep{Nicholl2018RNAAS...2..230N} to calculate the pseudo-bolometric luminosity and bolometric luminosity for SN~2021uvy using its ZTF $gri$ data and for SN~2022hgk using its ZTF $gri$ and ATLAS $co$ data. The other bands are first interpolated to $r$-band epochs, and then the pseudo-bolometric luminosity is calculated by integrating the fluxes over the bandpasses at each epoch. The bolometric luminosity is estimated from the pseudo-bolometric luminosity by adding blackbody corrections (absorbed UV and near-infrared). Figure~\ref{fig:lbol} shows the bolometric luminosity (top panel), estimated blackbody temperature (second panel), estimated blackbody radius (third panel), and $g-r$ color (bottom panel) for SNe~2019cad, 2019stc, 2021uvy, and 2022hgk, along with the regular Type Ic-BL SN~1998bw, and a slow-evolving single-peaked SLSN-I PS1-14bj (chosen for comparison as it also shows no color evolution during its decline). The data for SN~1998bw were obtained with \texttt{Superbol} using its UBVRI light curves \citep{Galama1998,Patat2001,Sollerman2002}. For SN~2019cad, only bolometric luminosity has been presented in \citet{Gutierrez2021} and not temperature or radius; therefore, we use the $groiz$ light curves from \citet{Gutierrez2021} and \texttt{Superbol} to calculate the shown data (we did not correct for host-extinction due to its high uncertainty). The data for SN~2019stc were obtained from \citet[][their fig.\,5]{Gomez2021} (they did not correct for host-extinction, although it is estimated in \citet{ChenYan2023} for SN~2019stc), and the data for PS1-14bj were obtained from \citet[their figs.\,7 and 8]{Lunnan2016}.

Figure~\ref{fig:lbol} shows that the first peaks of SN~2019stc and SN~2021uvy follow each other closely in bolometric luminosity, blackbody temperature, radius, and color. From the explosion until the end of the first decline (minima between the two peaks), both SNe show a consistent decrease in temperature (from $\sim10000$\,K to $\sim5000$\,K), an increase in radius, getting bluer during the rise and becoming redder during the first decline (which is typical of stripped-envelope supernovae powered by $^{56}$Ni, see SN~1998bw in gray). 
The similarity between SN~2019stc and SN~2021uvy stops at this point. For SN~2019stc, the temperature plateaus (like for SN~1998bw), and the radius follows the second brightening bump. However, for SN~2021uvy, the temperature starts rising rapidly along with no color evolution (like for PS1-14bj), staying around $g-r\approx0$ mag until very late times (indicating some new energy injection). At the same time, its radius declines at a similar rate as for PS1-14bj and SN~1998bw. This might indicate that the powering mechanisms of the second peaks of SN~2019stc and SN~2021uvy are different. For SN~2019stc, both radioactive decay and delayed magnetar engine are disfavored according to \citet{Gomez2021}, and an aspherical CSM, which could result in a lack of narrow lines, was instead favored for the second peak by those authors. We roughly estimate the $^{56}$Ni mass ($M_{Ni}$) and ejecta mass ($M_{ej}$) assuming that the first peaks of SNe~2019stc and 2021uvy are powered by radioactivity using the analytical expressions from \citet{KhatamiKasen2019ApJ}. This gives $M_{Ni}^{2019stc}\approx1.9$\,\Msun, $M_{Ni}^{2019uvy}\approx2.3$\,\Msun, $M_{ej}^{2019stc}\approx10$\,\Msun, and $M_{ej}^{2019uvy}\approx17$\,\Msun, which, as expected, are much too large compared to typical SESNe (and inconsistent with neutrino-driven core-collapse models), and thus make radioactivity as the only powering mechanism unfeasible.

\begin{table*}[htbp!]
    \caption{Peak bolometric and pseudo-bolometric luminosities and estimated radiated energies in the two peaks of our double-peaked SESN sample. The superscripts `1' and `2' denote the first and second peaks, respectively. SN~2021uvy has $\sim32 \times$ more energy radiated than SN~2022hgk (shown in bold).}
    \label{tab:lbol}
    \centering
    \begin{tabularx}{0.9\textwidth}{ccccccc}
    \toprule
    \toprule
       SN  & Lightcurve & $L_{pk}^{1}$ & $L_{pk}^{2}$ & $E_{rad}^{1}$ & $E_{rad}^{2}$ & $E_{rad}^{total}$  \\
         & & ($10^{43}$erg\,s$^{-1}$) & ($10^{43}$erg\,s$^{-1}$) & ($10^{50}$erg) & ($10^{50}$erg) & ($10^{50}$erg) \\
         \midrule
        2018ijp & Bolometric & $\sim1.5$ & $\sim0.6$  & $\sim0.3$  &  $\sim0.7$ & $\sim1.1$ \\
        2019cad & Bolometric & $0.13\pm0.02$ & $0.59\pm0.07$  & $0.008\pm0.001$  & $0.114\pm0.005$  &  $0.139\pm0.006$ \\
        2019oys & Bolometric & $>0.12$ & $0.10\pm0.39$ &  $>0.03$ & $0.405\pm0.296$ & $0.436\pm0.294$ \\
        2019stc & Bolometric & $\sim3.7$  & $\sim1.4$  & $\sim1.82$  & $\sim0.52$  & $\sim2.38$  \\
        2020acct & Bolometric & $0.96\pm0.06$  & $0.35\pm0.01$  & $\sim0.07$  & $\sim0.04$  & $\sim0.15$  \\
        2021pkd & Bolometric  & $0.85\pm1.13$  &  $0.34\pm0.09$ &  $0.092\pm0.018$ & $0.059\pm0.007$  & $0.154\pm0.019$ \\
       \textbf{2021uvy} & Pseudo-bolometric & $1.08 \pm 0.03$ & $0.72 \pm 0.04$ & $0.371 \pm 0.004$ & $0.764 \pm 0.015$ & $1.160 \pm 0.016$ \\
        & Bolometric & $3.88\pm0.77$ & $2.30\pm0.76$ & $1.070\pm0.037$ & $2.244\pm0.181$ & \textbf{3.367$\pm$0.183} \\
       \textbf{2022hgk} & Pseudo-bolometric & $0.05\pm0.01$ & $0.14\pm0.01$ & $0.007\pm0.001$ & $0.037\pm0.001$ & $0.045\pm0.001$ \\  
       &  Bolometric & $0.18\pm0.05$ & $0.35\pm0.10$ & $0.021\pm 0.001$ & $0.095\pm0.006$ & \textbf{0.117$\pm$0.006} \\
        2022jli & Pseudo-bolometric & $\sim0.3$  & $\sim0.4$  &  $\sim0.05$ & $\sim0.29$ & $\sim0.35$  \\
        2022xxf & Bolometric & $\sim0.9$  & $\sim1.3$  &  $\sim0.22$ & $\sim0.42$  &  $\sim0.67$ \\
        2023aew & Bolometric & $0.07\pm0.00$  & $1.20\pm0.20$  & $0.096\pm0.005$  & $0.560\pm0.013$  &  $0.656\pm0.018$ \\
        2023plg & Bolometric  & $>0.19$  & $0.67\pm0.04$  &  $>0.04$ & $0.218\pm0.004$  & $0.258\pm0.008$ \\
       \bottomrule
    \end{tabularx}
\end{table*}

On the other hand, SN~2022hgk's bolometric light curve almost exactly matches that of SN~2019cad (if not corrected for host extinction), except towards the very end, when SN~2022hgk shows a little bump before fading completely. The temperature mirrors the luminosity and decreases sharply during the first decline (same as SNe~1998bw, 2019cad, 2019stc, 2021uvy), shows a small rise during the second brightening (like SN~2019cad, SN~2021uvy), decreases again during the second decline (like SN~2019cad), and rises at the very end (coincident with the final luminosity bump). SN~2022hgk's radius only shows a rise and a decline, peaking around the second (and brightest) luminosity maximum. SN~2022hgk's $g-r$ color becomes progressively redder during the second decline, as expected, and has a similar evolution to the $g-r$ color of SN~2019cad.

In Table~\ref{tab:lbol}, we have collected bolometric (and in some cases pseudo-bolometric when the bolometric estimate is not provided) luminosities at the two light-curve peaks and the estimated total radiated energies ($E_{rad}$) to crudely compare the energetics across the sample. We integrate bolometric light curves of SNe~2019cad, 2019oys, 2021pkd, 2021uvy, 2022hgk, and 2023plg obtained using \texttt{Superbol} and ZTF light curves to estimate the radiated energies and use the Monte-Carlo method to estimate the uncertainties on radiated energies as follows. We sample 1000 random points per epoch from a normal distribution that has the epoch luminosity as the mean and the uncertainty on the luminosity as the $\sigma$. We integrate the sampled light curves over the rest-frame days and take the mean and standard deviation of the resulting energy estimates. For SN~2023aew, we list the values reported in \citet{Sharma2024} that have been estimated using the same process described above. For SNe~2018ijp, 2019stc, 2020acct, 2022jli, and 2022xxf we integrate the bolometric (or pseudo-bolometric) light curves obtained from \citet[their fig.\,2]{Tartaglia2021}, \citet[their fig.\,5]{Gomez2021}, \citet[their fig.\,9]{Angus2024}, \citet[fig.\,4]{Chen2024Natur.625..253C}, and \citet[their fig.\,A.1]{Kuncarayakti2023} respectively. We simply consider points from the first detection to the local minimum between the two peaks for calculating the energy radiated in the first peak and from the local minimum to the last detection for calculating the energy radiated in the second peak. This provides the simplest lower limits for the radiated energies, as we are not fitting any specific powering mechanisms to the light curves. 

\subsection{Spectral comparison}

\begin{figure*}[htbp!]
    \centering
    \includegraphics[width=0.98\textwidth]{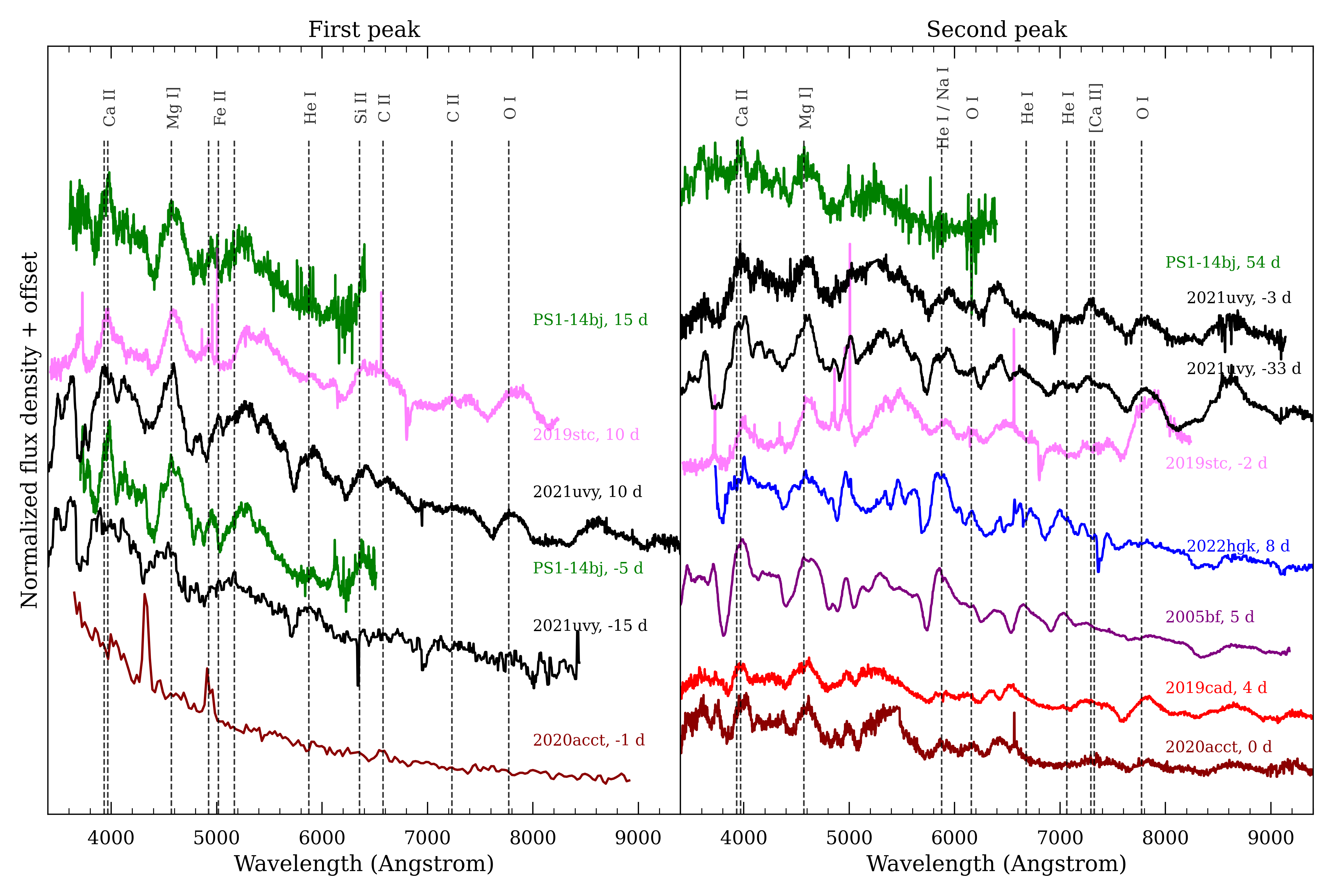}
    \caption{\textit{Left} First-peak spectra of SN~2021uvy (black) compared with those of SNe~2019stc (magenta), 2020acct (brown), and PS1-14bj (green), with phases reported with respect to the first peak. Similar to normal SESNe, the first-peak spectra of SNe~2019stc, 2021uvy and PS1-14bj are dominated by \ion{Ca}{2}, \ion{Mg}{1}], \ion{Fe}{2} and \ion{O}{1}. SN~2020acct, on the other hand, shows signs of CSM interaction at this phase. \textit{Right} Second-peak spectra of SNe~2021uvy (black) and 2022hgk (blue) compared with those of SNe~2005bf (purple), 2019cad (red), 2019stc (magenta), 2020acct (brown), and PS1-14bj (green), with phases reported with respect to the second peak (except for PS1-14bj). SN~2022hgk shows a close spectroscopic resemblance to the peculiar Type~Ib SN~2005bf around their main (second) peaks. All spectra are smoothed with a median filter of window size 5 (except for SN~2020acct).}
    \label{fig:speccomp1}
\end{figure*}

Figure~\ref{fig:speccomp1} compares the spectra obtained near the first (left panel) and second (right panel) peaks of SNe~2021uvy and 2022hgk with the most similar double-peaked SESNe from the sample. The first-peak spectra of SN~2021uvy have normal SESN features and look similar to those of SN~2019stc and PS1-14bj. SN~2021uvy shows \ion{He}{1} $\lambda5876$ signatures from the pre-peak epochs (Figure~\ref{fig:specseq}, left) which classifies it as a Type~Ib. From the absorption minima of \ion{O}{1} $\lambda7774$ in the day 10 spectrum, we estimate an ejecta velocity of $\sim8000$\,\kms, which is also consistent with the \ion{He}{1} absorption minimum. The lines of \ion{Ca}{2} $\lambda\lambda 3934, 3969$, \ion{Mg}{1}] $\lambda4571$, and \ion{O}{1} $\lambda7774$ appear to be of similar strength in these three SNe. The \ion{Fe}{2} complex between 5000\,\AA\ and 5600\,\AA\ has more flux on the blue side and appears broader in SN~2021uvy than for SN~2019stc and PS1-14bj. SN~2021uvy and PS1-14bj also appear to have a slightly bluer continuum than SN~2019stc past the first peak. Overall, SN~2021uvy's first peak exhibits Type~Ib nature spectrally but with a slow-evolving SLSN-like light curve that hints towards a mixed powering mechanism (radioactivity $+$ magnetar) as suggested by \citet{Gomez2021,Gomez2022}.

Other double-peaked SESNe that exhibit normal SESN spectra during the first peak include SNe~2019cad, 2022jli, and 2022xxf. However, SN~2022jli evolved into having accretion-powered second peak \citep{Chen2024Natur.625..253C}, and SN~2022xxf developed subtle H/He-free signs of CSM interaction \citep{Kuncarayakti2023}. SN~2023aew changed its type from SN~II during the first peak to SN~Ic during its second peak and then to having hydrogen reappear during the nebular phase, which could be due to hidden CSM interaction with a complex geometry \citep{Sharma2024}. SN~2020acct showed some early flash-ionization features, a sign of brief CSM interaction during the first peak \citep{Angus2024}, confirming its power source. This is to say that the sample of double-peaked SESNe show as much variety in their spectral nature as they do in their light curves and intermediate resolution spectra taken at crucial epochs in the light-curve evolution (early rise, peak, minima between peaks, second peak, and nebular) are necessary to enable the identification of the powering mechanism. Unfortunately, for SN~2022hgk, no first-peak spectra were taken as it remained below the threshold for triggering follow-up as part of the BTS survey.

\begin{figure*}[htbp!]
    \centering
    \includegraphics[width=0.98\textwidth]{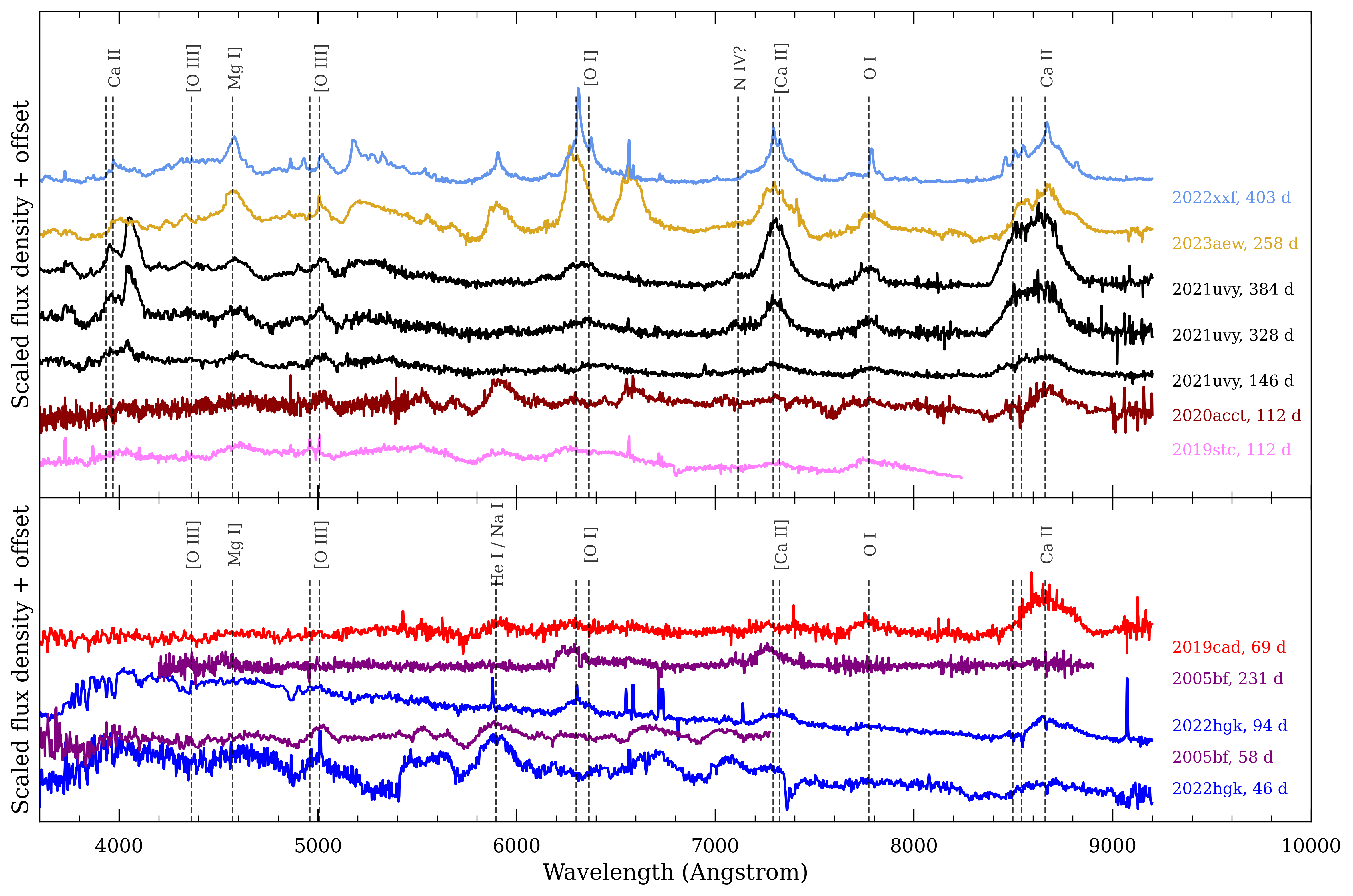}
    \caption{\textit{Top} Comparison of nebular spectra of SN~2021uvy (black) with SNe~2019stc (magenta), 2020acct (brown), 2022xxf (cornflowerblue) and 2023aew (gold). \textit{Bottom} Comparison of nebular spectra of SN~2022hgk (blue) with SNe~2005bf (purple) and 2019cad (red). All spectral phases are reported with respect to the first peak.}
    \label{fig:specneb}
\end{figure*}

Looking at the right panel of Figure~\ref{fig:speccomp1}, around the second peak, the broad features of SNe~2019cad, 2019stc, 2020acct, and 2021uvy are similar and post-peak SESN-like but redder than PS1-14bj. The Ca NIR bumps also become prominent in SNe~2019cad, 2020acct, and 2021uvy. SN~2021uvy also has emission around 7300\,\AA\ which could possibly be [\ion{Ca}{2}] $+$ [\ion{O}{2}], which is unusual for typical SESNe in photospheric phase, but has been observed in SNe~2019stc, 2020acct, 2023aew, and also 2018ibb \citep{Schulze2024} as noted in \citet{Angus2024}. \citet{Angus2024} also noted the striking similarity of the second-peak spectra of SNe~2020acct and 2023aew. SN~2022hgk shows strong \ion{He}{1} lines at this epoch ($\sim10000$\,\kms) and a blue continuum. The SN~2022hgk spectrum at 7 days after the second peak closely resembles SN~2005bf's spectrum at 5 days past the second peak \citep{Shivvers2019}, with both showing \ion{He}{1} lines and a lack of \ion{O}{1} $\lambda 7774$. None of these SNe show obvious signs of interaction in their second peak spectra.

Figure~\ref{fig:specneb} shows nebular (and near-nebular) spectra of some double-peaked SESNe, with the common nebular lines marked and some tentative line identifications. The phases shown are from the estimated time of the explosion. 
The final spectra available of SNe~2019stc and 2020acct are shown in the top panel and though they are not fully nebular, we can see [\ion{O}{1}] $\lambda\lambda6300, 6364$ and [\ion{Ca}{2}] $\lambda\lambda7292, 7324$ starting to appear. The spectra of SNe~2022xxf, 2023aew, and 2021uvy in the top panel have slight differences that could allude to their origin. Narrow lines become discernible in the nebular spectra of SN~2022xxf, revealing the H/He-free CSM interaction. SN~2023aew shows strong emission at the location of H$\alpha$, which appears to be too strong to be the [\ion{N}{2}] nebular emission seen in many Type~IIb/Ib \citep{Sharma2024,Barmentloo2024} and instead could be re-emerged H$\alpha$, revealing the hidden CSM powering the supernova. However, the [\ion{Ca}{2}]/[\ion{O}{1}] flux ratio in these SNe (2021uvy $\sim1.18$, 2022xxf $\sim1.16$, 2023aew $\sim0.8$, 2022hgk $\sim0.92$, 2005bf $\sim0.90$) are similar, indicating similar oxygen core masses and in turn similar progenitors. SN~2021uvy shows strong emission lines around $\sim4000$\,\AA\ which could be \ion{Ca}{2} H\&K lines but appear to be redshifted. The [\ion{Ca}{2}] line in SN~2021uvy maintains a Gaussian profile with time, but [\ion{O}{1}] seems to become flat-topped (similar to the case of Type Ib iPTF13bvn; \citealp{Kuncarayakti2015}), especially in the 384-day spectrum. This could be due to some asphericity in the ejecta (clumps or torus-like oxygen distribution as suggested in \citealt{Taubenberger2009MNRAS}), or it could be due to absorption in the interior \citep{Milisavljevic2010ApJ}.

The bottom panel of Figure~\ref{fig:specneb} compares SNe~2019cad and 2022hgk with SN~2005bf. The 46, 58, and 69-day spectra of SNe~2022hgk, 2005bf, and 2019cad, respectively, show hints of nebular emission lines but are not fully nebular. SN~2019cad differs from the other two SNe and shows stronger \ion{O}{1} emission. SN~2022hgk at 46 days matches SN~2005bf at 58 days, maintaining the spectral similarities since their peaks. The 231-day spectrum of SN~2005bf shows its characteristic blueshifted nebular lines, but the 94-day spectrum of SN~2022hgk does not, which is where SN~2022hgk finally differs from SN~2005bf. The blue continuum in SN~2022hgk at this phase is likely contamination from the host galaxy. 

\section{Discussion}\label{sec:disc}

\subsection{Trends in the double-peaked light curve properties}\label{sec:lcparams}

\begin{figure*}[htbp!]
    \centering
    \includegraphics[width=0.95\textwidth]{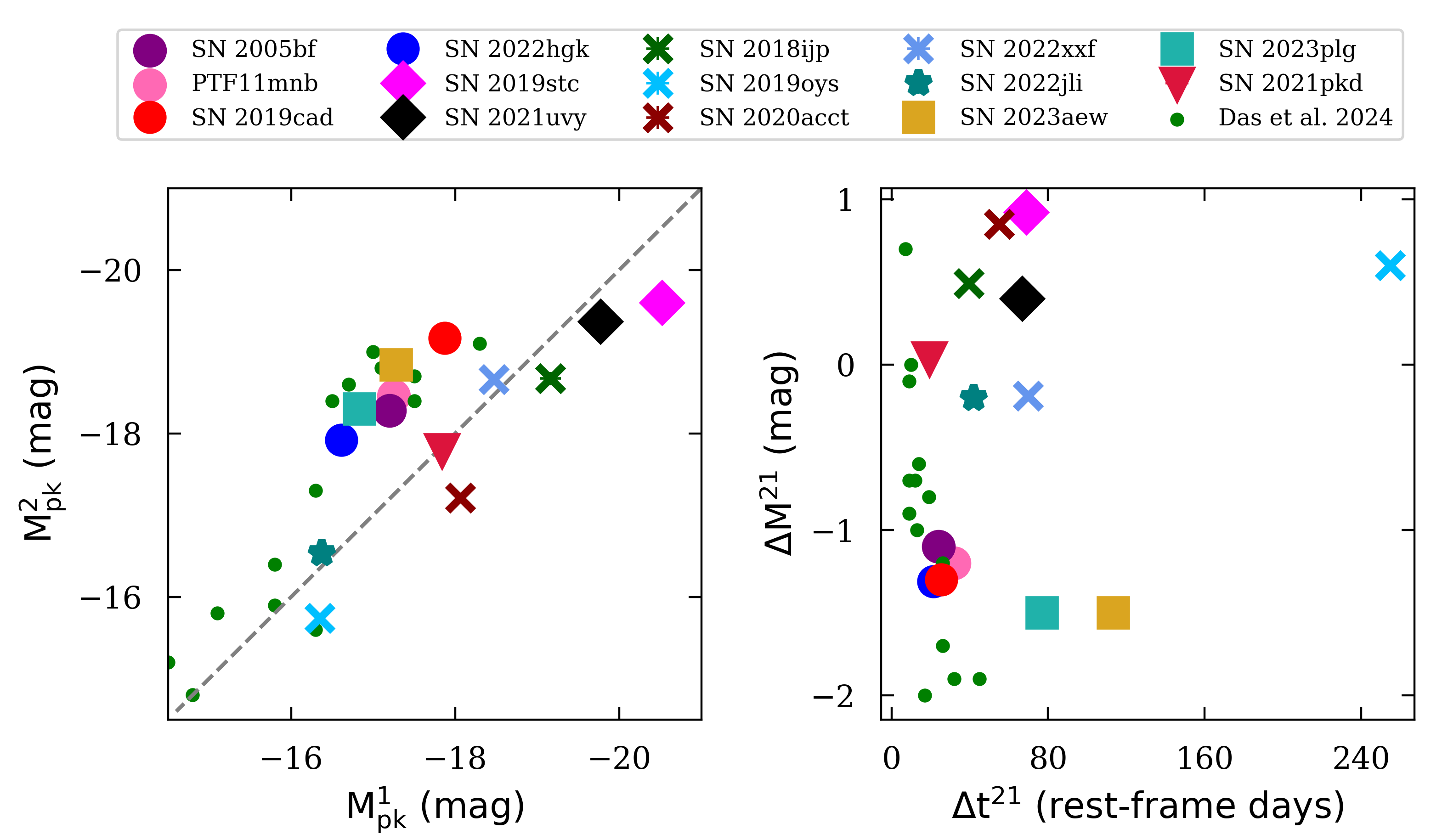}
    \caption{\textit{Left} Peak absolute magnitudes of the second peak vs. the first peak for the double-peaked SESN sample, and the shock-cooling powered double-peaked SESNe presented in \citet{Das2024}. There appears to be a correlation between the peak magnitudes, which is strongest for the \citet{Das2024} SESNe (p-value $<10^{-5}$) but also significant for our double-peaked SESN sample (p-value $=0.005$). \textit{Right} Magnitude difference vs. rest-frame duration between the two peaks. Again, the potentially double-nickel powered SESNe form a tight group in this phase space.}
    \label{fig:m1m2}
\end{figure*}

The left panel of Figure~\ref{fig:m1m2} shows the $r$-band ($g$-band for SN~2022jli) absolute magnitudes of the second peak against those of the first peak ($M^2_{pk}$ vs $M^1_{pk}$), the right panel depicts the difference between the peak magnitudes against rest-frame duration between the peaks ($\Delta M^{21}$ vs $\Delta t^{21}$) for the double-peaked SESN sample discussed in this paper and for the sample of double-peaked Type~Ibc SNe presented in \citet{Das2024}. There appears to be a correlation between the absolute magnitudes of the first and second peaks (p-value $=0.005$). The absolute magnitude correlation was observed by \citet{Das2024} for double-peaked SESNe that have the first peak attributed to cooling after the shock passes the extended envelope (or nearby CSM) of the progenitor. The mechanism behind such a correlation remains unclear. One possibility that \citet{Das2024} put forth is that SESNe with shock-cooling first peaks have He-star progenitors that shed their envelopes in binary interactions shortly before exploding. For such progenitors, the first peak depends on the progenitor radius and the second peak on the $^{56}$Ni mass. In both panels of Figure~\ref{fig:m1m2}, SNe~2005bf, PTF11mnb, 2019cad, and 2022hgk (all potentially powered by double-nickel distributions, marked with circles) seem to form a group and lie in the same phase-space as the SESNe with shock-cooling peaks. The correlation in shock-cooling powered and double-nickel powered cases could also stem from both peaks being positively correlated with the explosion energy. SESNe with at least one of the peaks potentially powered by CSM interaction (marked with crosses) and the accretion-powered SN~2022jli (marked with a star) follow the correlation in the left panel but do not seem to form a group. Finally, SNe~2019stc and 2021uvy form a close duo in all panels.

It is apparent from Figure~\ref{fig:m1m2} that the location these supernovae occupy in the different phase spaces created by the light curve properties ($M^2_{pk}-M^1_{pk}$, $\Delta M^{21}-\Delta t^{21}$, etc.) can 
help unveil the possible powering mechanisms, especially for models that have quantifiable restrictions on these light curve properties.

\subsection{Powering Mechanisms}\label{sec:powering}

The double-peaked stripped-envelope supernovae discussed so far exemplify the uncertainty about the powering mechanism of the light curves of this class. Normal SNe~Ibc can be relatively well explained as being powered by the decay of radioactive $^{56}$Ni that diffuses out of the initially optically thick ejecta. This self-contained explanation follows the simple model by \cite{Arnett1982ApJ...253..785A}. It should be mentioned, however, that even this picture has been questioned in the literature. The ejecta masses deduced from some light curve analysis studies indicate values lower than anticipated from massive single stars \citep[e.g.,][]{Taddia2015A&A...574A..60T,Prentice2019MNRAS.485.1559P} and the $^{56}$Ni masses are too high to be explained by contemporary neutrino-driven core collapse models \citep{Sollerman2022A&A...657A..64S}, spurring discussion on the need for other powering mechanisms even for the normal objects \citep[e.g.,][]{Rodriquez2024Natur.628..733R,Karamehmetoglu2023A&A...678A..87K}. Analysis of the relationship between nebular line flux ratios ([\ion{Ca}{2}]/[\ion{O}{1}]) and ejecta masses estimated from light curve modeling (with \citealt{Arnett1982ApJ...253..785A}) of SESNe also revealed no connection between the two, meaning both low and high ejecta mass objects have similar progenitors, implying the presence of other powering mechanisms responsible for the light-curve behavior \citep{2022MNRAS.514.5686P}. Studying the rarer family of double-peaked objects has provided a plethora of suggestions, including the most common scenarios for powering the emission of supernovae. Often, different mechanisms or a combination thereof are invoked to explain each peak in such supernovae, although some modeling studies exist that try to explain the double-peaked light curve with a single mechanism. In the following sections, we briefly discuss the suggested powering mechanisms and attempt to form a picture of their diversity.

\subsubsection{Double-Nickel distribution}

An early suggestion for double-peaked SESNe was the notion of double nickel distributions. A jet-like structure that brings some radioactive material closer to the surface was proposed for the double-peaked peculiar Type~Ib SN~2005bf \citep{Folatelli2006}, which would produce an early light curve peak before the central Ni power diffuses out on a longer time scale. SN~2019cad (analog of SN~2005bf) was proposed to have such a structure \citep[][see also PTF11mnb]{Gutierrez2021,Taddia2018}, and the scenario was further explored \cite[e.g.,][]{Orellana2022A&A...667A..92O}. This mechanism fits well given SN~2022hgk's striking photometric similarity with SN~2019cad and spectral similarity with SN~2005bf. 
It is clear from these models, however, that they have limited ability to match light-curve peaks that are well separated (large $\Delta t^{21}$ and in turn delayed second peak that would be inconsistent with radioactive power diffusion timescale, like SNe~2019stc, 2020acct, 2021uvy, 2023aew), or with high luminosities (that require unreasonable nickel mass like SN~2019stc and SN~2021uvy), or that have more than two peaks, and the model is thus not generic enough to explain the full double peaked sample of SESNe.

\subsubsection{Magnetar}
\label{sec:magnetar}

The magnetar model has become popular for long-lived transients where the Arnett model yields unphysical $^{56}$Ni masses, and is often invoked for peculiar SESNe (like SN~2005bf; \citealp{Maeda2007}), luminous SNe \citep{Gomez2022} and superluminous supernovae \cite[SLSNe, e.g.,][]{ChenSLSNZTFa2023ApJ...943...42C}. The model offers a lot of flexibility in terms of rise times, peak luminosities, and duration -- but does not naturally allow for double-peaked light curves or undulations. \citet{Chugaimagnetar2022MNRAS.512L..71C} opposed the CSM-interaction scenario for the second peak of the luminous SN~2019stc as put forth in \citet{Gomez2021}, and instead suggested a magnetar engine by invoking a less-understood dipole-field enhancement to allow for the second peak. Other similar suggestions, like magnetar flare activity, have been proposed in the context of wiggly light curves of SLSNe \citep{Dong2023ApJ...951...61D,Zhu2024ApJ...970L..42Z}, and \citet{MoriyaMagnetar2022MNRAS.513.6210M} suggested that variations in the thermal energy injection from magnetar spin-down cause the light-curve bumps. However, \citet{Chugaimagnetar2022MNRAS.512L..71C} only explains a single bump and does not identify any specific smoking-gun observables that could support the model. \citet{MoriyaMagnetar2022MNRAS.513.6210M}, on the other hand, predicts an increase in photospheric temperature coincident with the bumps and notes that SN~2019stc does not show such an increase. The only supernova in our sample that shows an increase in photospheric temperature for the second peak is SN~2021uvy, and therefore, could be an example of the magnetar thermal energy injection scenario. However, our temperature measurements only use $gri$ bands, but UV data are required for a more accurate temperature estimate, and thus, this observation of temperature rise is tentative. The temperature rise in SN~2021uvy also appears to last throughout the entire duration of the second peak, implying that the increase in thermal energy injection would also need to be maintained for $>100$ days. 

\subsubsection{CSM}

While some double-peaked SESNe have shown strong signs of interaction after the first peak that completely transform their spectra -- for example, hydrogen-rich CSM interaction in SNe~2018ijp \citep{Tartaglia2021} and 2019oys \citep{Sollerman2020}; others have shown much more subtle but revealing signs of CSM interaction. One example of such a case is SN 2022xxf \citep{Kuncarayakti2023}, where the evidence for CSM interaction became obvious only at later times when narrow emission lines became more apparent in the optical spectra. The CSM must, in this case, be poor in both hydrogen and helium, which makes the configuration highly unusual (a detached CSM model was suggested for SN~2022xxf by \citealt{TakeiTsuna2024}). The analytical modeling by \citet{Chiba2024ApJ...973...14C} explicitly mentions the possibility of modeling both of the peaks in the light curves of SNe~2005bf and 2022xxf using a flat density profile for the CSM. However, the model comes with the caveat that the duration between the two peaks ($\Delta t^{21}$) can be at most $\lesssim100$ days, otherwise, the ejecta mass requirements become unphysical. Another caveat is that if the two peaks are too temporally separated (large $\Delta t^{21}$), the breakout luminosity (first peak) cannot be comparable to the luminosity of the second peak and thus the model has difficulty in explaining cases where first peak is brighter than the second peak (e.g., SN~2019stc). \citet{KhatamiKasen2024ApJ...972..140K} explore different theoretical scenarios enabling a large variety of light curves from the CSM interaction powering only, including double-peaked light curves which in their modeling occur when the shock breaks out just outside the CSM edge (so that there is no continued interaction phase, see \citealp[their fig.\,3]{KhatamiKasen2024ApJ...972..140K}) and the CSM is ``heavy" (CSM mass $\gtrsim$ ejecta mass, making the shock cooling phase more prominent). However,  spectral signatures of such heavy CSM might be difficult to hide, thus making this scenario less likely for SESNe without any narrow line signatures. In the case of SN~2023aew \citep{Kangas2024,Sharma2024}, the H$\alpha$ P-Cygni feature seen during the first peak vanished at the time of the second peak and appeared again at later times, and the nebular lines showed a ``horned" structure. These features, combined with the double-peaked light curve with large $\Delta t^{21}$, could be evidence that an aspherical or clumpy CSM powers the second peak of the supernova along with radioactive nickel decay, with the first peak being an eruptive precursor. SN~2023plg \citep{Sharma2024} follows the light-curve behavior of SN~2023aew, and its second peak spectra share strong similarities with SN~2023aew's second peak spectra \citep[their fig.\,10]{Sharma2024}, and could share the same powering mechanism.

\subsubsection{Accretion}

Another potential powering mechanism is accretion onto a compact object, where an accretion disk might form and efficiently convert energy to radiation. SN~2022jli \citep{Chen2024Natur.625..253C} -- the double-peaked SESN showing periodic undulations in its light curve during the second decline, was potentially powered by such a scenario. \cite{Chen2024Natur.625..253C} advocated that the first peak might have been powered by a normal radioactive decay, whereas the second peak would be powered by mass accretion from the companion onto the newly formed compact object remnant. The second peak of this supernova was instead suggested to be powered by a magnetar (\S\ref{sec:magnetar}) by \cite{Cartier2024arXiv241021381C}. In general, the different powering scenarios mentioned in these sections have been combined in a variety of different ways to explain double-peaked SESNe.

\subsubsection{Pulsational Pair Instability mechanism}

Finally, we mention the suggestion put forward by \cite{Angus2024} for SN~2020acct, that the double-peaked light curve could have been powered by CSM interaction with a configuration from a pulsational pair instability supernova (PPISN; \citealp{Barkat1967,Rakavy1967}). PPI events cause extreme mass loss, and thus their ejecta CSM interactions can be quite luminous. The timing of the different events can vary depending on the specific evolution of the system and therefore provide models that can fit multiple well-separated peaks, explain precursors, and also bumpy light curves \citep{2017ApJ...836..244W,2019ApJ...887...72L}. However, clear identification of PPISNe is difficult as other powering mechanisms (and their combinations) could also fit the observations of peculiar multi-peaked SNe, and the surrounding CSM could also come from various mass-loss mechanisms (LBV eruptions, winds, etc.). The unique properties of SN~2020acct -- hydrogen-poor interaction signatures during the first peak and a second peak showing terminal explosion SESN-like properties, together with an unfeasible nickel fraction ($f_{Ni}\sim0.91$) from fitting radioactive decay power to the second peak, made it a possible PPISN candidate \citep{Angus2024}.

\section{Summary}\label{sec:summary}

In this paper, we have presented optical photometry and spectroscopy of two double-peaked stripped envelope supernovae discovered by the Zwicky Transient Facility. We discuss the comprehensive dataset in conjunction with a sample of previously reported, clearly double-peaked stripped-envelope supernovae from the ZTF archive, and for several of these, we also provide previously unpublished data. With data from one homogeneous survey, we can quantify some of the key properties of the double-peaked light curves, analyze correlations between these properties, and contextualize them with some of the common powering mechanisms that we review from the literature.

SN~2021uvy is a luminous and slowly evolving Type~Ib supernova with both peaks reaching roughly the same brightness. Although it shows many similarities to SN~2019stc, with both having their first peaks fitting a combination of radioactive nickel power and magnetar central engine input, their second peaks diverge significantly in behavior. SN~2021uvy shows a lack of color evolution during the second decline and a rise in photospheric temperature, which is a prediction in the case of variable thermal energy injection from magnetar spin-down \citep{MoriyaMagnetar2022MNRAS.513.6210M}.

SN~2022hgk, on the other hand, is an average-luminosity Type~Ib supernova with a much brighter second peak. Its light curve is very similar to the light curve of  SN~2019cad, which is considered an analog of SN~2005bf (and also to PTF11mnb). The spectra of SN 2022hgk, however, show a significant similarity with those of  SN~2005bf (strong helium absorption features) rather than with those of SN~2019cad. Overall, these four supernovae (SNe~2005bf, PTF11mnb, 2019cad, and 2022hgk) have similar light-curve parameters and form a tight group in the phase space of absolute peak magnitudes of the second peak vs. that of the first peak and in the magnitude difference between the peaks vs. the duration between the peaks. The double-nickel distribution powering mechanism might well fit this group of supernovae \citep[see e.g.,][]{Orellana2022A&A...667A..92O}.

With a sample of double-peaked SESNe coming together (being $\sim2.5\%$ of all Type~Ibc SNe), it becomes clear that this is a phenomenon that requires a more holistic approach. There have been good arguments in the literature as to why some of these events should not be just random alignments of two distinct SNe, or even two separate stars exploding in a binary system, and with the expanding sample, such probability estimates gain more weight. At the same time, fine-tuned models to explain individual and very rare systems become less probable once it is realized that more of these systems exist. Upcoming facilities like the Rubin Observatory will increase the sample size of double-peaked and multi-peaked SESNe and also provide more light curve properties to help uncover their powering mechanisms with the depth of the Legacy Survey of Space and Time (LSST; \citealp{lsst}).

\section{Acknowledgement}

\small{Based on observations obtained with the Samuel Oschin Telescope 48-inch and the 60-inch Telescope at the Palomar Observatory as part of the Zwicky Transient Facility project. ZTF is supported by the National Science Foundation under Grants No. AST-1440341 and AST-2034437, and a collaboration including current partners Caltech, IPAC, the Oskar Klein Center at Stockholm University, the University of Maryland, University of California, Berkeley, the University of Wisconsin at Milwaukee, University of Warwick, Ruhr University Bochum, Cornell University, Northwestern University, and Drexel University. Operations are conducted by COO, IPAC, and UW. 
The ZTF forced-photometry service was funded under the Heising-Simons Foundation grant \#12540303 (PI: Graham). 
SED Machine is based upon work supported by the National Science Foundation under Grant No. 1106171. 
This work was supported by the GROWTH project \citep{Kasliwal_2019} funded by the National Science Foundation under PIRE Grant No 1545949. 
The Gordon and Betty Moore Foundation, through both the Data-Driven Investigator Program and a dedicated grant, provided critical funding for SkyPortal. 
The Oskar Klein Centre was funded by the Swedish Research Council. Partially based on observations made with the Nordic Optical Telescope, operated by the Nordic Optical Telescope Scientific Association at the Observatorio del Roque de los Muchachos, La Palma, Spain, of the Instituto de Astrofisica de Canarias. Some of the data presented here were obtained with ALFOSC, which is provided by the Instituto de Astrofisica de Andalucia (IAA) under a joint agreement with the University of Copenhagen and NOT. 
Some of the data presented herein were obtained at the W. M. Keck Observatory, which is operated as a scientific partnership among the California Institute of Technology, the University of California, and NASA; the observatory was made possible by the generous financial support of the W. M. Keck Foundation. 
This work has made use of data from the Asteroid Terrestrial-impact Last Alert System (ATLAS) project. The Asteroid Terrestrial-impact Last Alert System (ATLAS) project is primarily funded to search for near earth asteroids through NASA grants NN12AR55G, 80NSSC18K0284, and 80NSSC18K1575; byproducts of the NEO search include images and catalogs from the survey area. The ATLAS science products have been made possible through the contributions of the University of Hawaii Institute for Astronomy, the Queen’s University Belfast, the Space Telescope Science Institute, the South African Astronomical Observatory, and The Millennium Institute of Astrophysics (MAS), Chile. 
This research has made use of the NASA/IPAC Infrared Science Archive, which is funded by the National Aeronautics and Space Administration and operated by the California Institute of Technology. 
The Liverpool Telescope is operated on the island of La Palma by Liverpool John Moores University in the Spanish Observatorio del Roque de los Muchachos of the Instituto de Astrofisica de Canarias with financial support from the UK Science and Technology Facilities Council. 
 
Y. Sharma thanks the LSSTC Data Science Fellowship Program, which is funded by LSSTC, NSF Cybertraining Grant \#1829740, the Brinson Foundation, and the Moore Foundation; her participation in the program has benefited this work.

M.W.C acknowledges support from the National Science Foundation with grant numbers PHY-2308862 and PHY-2117997.
 
\texttt{Fritz} \citep{skyportal2019,Coughlin2023} and GROWTH marshal \citep{Kasliwal_2019} (dynamic collaborative platforms for time-domain astronomy) were used in this work.}

\bibliography{biblio}

\begin{thebibliography}{}
\expandafter\ifx\csname natexlab\endcsname\relax\def\natexlab#1{#1}\fi
\providecommand{\url}[1]{\href{#1}{#1}}
\providecommand{\dodoi}[1]{doi:~\href{http://doi.org/#1}{\nolinkurl{#1}}}
\providecommand{\doeprint}[1]{\href{http://ascl.net/#1}{\nolinkurl{http://ascl.net/#1}}}
\providecommand{\doarXiv}[1]{\href{https://arxiv.org/abs/#1}{\nolinkurl{https://arxiv.org/abs/#1}}}

\bibitem[{{Angus} {et~al.}(2024){Angus}, {Woosley}, {Foley}, {Nicholl}, {Villar}, {Taggart}, {Pursiainen}, {Ramsden}, {Srivastav}, {Stevance}, {Moore}, {Auchettl}, {Hoogendam}, {Khetan}, {Yadavalli}, {Dimitriadis}, {Gagliano}, {Siebert}, {Aamer}, {Boer}, {Chambers}, {Clocchiatti}, {Coulter}, {Drout}, {Farias}, {Fulton}, {Gall}, {Gao}, {Izzo}, {Jones}, {Lin}, {Magnier}, {Narayan}, {Ramirez-Ruiz}, {Ransome}, {Rest}, {Smartt}, \& {Smith}}]{Angus2024}
{Angus}, C.~R., {Woosley}, S.~E., {Foley}, R.~J., {et~al.} 2024, \apjl, 977, L41, \dodoi{10.3847/2041-8213/ad9264}

\bibitem[{{Anupama} {et~al.}(2005){Anupama}, {Sahu}, {Deng}, {Nomoto}, {Tominaga}, {Tanaka}, {Mazzali}, \& {Prabhu}}]{Anupama2005}
{Anupama}, G.~C., {Sahu}, D.~K., {Deng}, J., {et~al.} 2005, \apjl, 631, L125, \dodoi{10.1086/497336}

\bibitem[{{Arnett}(1982)}]{Arnett1982ApJ...253..785A}
{Arnett}, W.~D. 1982, \apj, 253, 785, \dodoi{10.1086/159681}

\bibitem[{{Astropy Collaboration} {et~al.}(2022){Astropy Collaboration}, {Price-Whelan}, {Lim}, {Earl}, {Starkman}, {Bradley}, {Shupe}, {Patil}, {Corrales}, {Brasseur}, {N{\"o}the}, {Donath}, {Tollerud}, {Morris}, {Ginsburg}, {Vaher}, {Weaver}, {Tocknell}, {Jamieson}, {van Kerkwijk}, {Robitaille}, {Merry}, {Bachetti}, {G{\"u}nther}, {Aldcroft}, {Alvarado-Montes}, {Archibald}, {B{\'o}di}, {Bapat}, {Barentsen}, {Baz{\'a}n}, {Biswas}, {Boquien}, {Burke}, {Cara}, {Cara}, {Conroy}, {Conseil}, {Craig}, {Cross}, {Cruz}, {D'Eugenio}, {Dencheva}, {Devillepoix}, {Dietrich}, {Eigenbrot}, {Erben}, {Ferreira}, {Foreman-Mackey}, {Fox}, {Freij}, {Garg}, {Geda}, {Glattly}, {Gondhalekar}, {Gordon}, {Grant}, {Greenfield}, {Groener}, {Guest}, {Gurovich}, {Handberg}, {Hart}, {Hatfield-Dodds}, {Homeier}, {Hosseinzadeh}, {Jenness}, {Jones}, {Joseph}, {Kalmbach}, {Karamehmetoglu}, {Ka{\l}uszy{\'n}ski}, {Kelley}, {Kern}, {Kerzendorf}, {Koch}, {Kulumani}, {Lee}, {Ly}, {Ma}, {MacBride}, {Maljaars}, {Muna}, {Murphy}, {Norman},
  {O'Steen}, {Oman}, {Pacifici}, {Pascual}, {Pascual-Granado}, {Patil}, {Perren}, {Pickering}, {Rastogi}, {Roulston}, {Ryan}, {Rykoff}, {Sabater}, {Sakurikar}, {Salgado}, {Sanghi}, {Saunders}, {Savchenko}, {Schwardt}, {Seifert-Eckert}, {Shih}, {Jain}, {Shukla}, {Sick}, {Simpson}, {Singanamalla}, {Singer}, {Singhal}, {Sinha}, {Sip{\H{o}}cz}, {Spitler}, {Stansby}, {Streicher}, {{\v{S}}umak}, {Swinbank}, {Taranu}, {Tewary}, {Tremblay}, {de Val-Borro}, {Van Kooten}, {Vasovi{\'c}}, {Verma}, {de Miranda Cardoso}, {Williams}, {Wilson}, {Winkel}, {Wood-Vasey}, {Xue}, {Yoachim}, {Zhang}, {Zonca}, \& {Astropy Project Contributors}}]{astropy2022}
{Astropy Collaboration}, {Price-Whelan}, A.~M., {Lim}, P.~L., {et~al.} 2022, \apj, 935, 167, \dodoi{10.3847/1538-4357/ac7c74}

\bibitem[{Barbary(2016)}]{extinction2016}
Barbary, K. 2016, extinction v0.3.0,  Zenodo, \dodoi{10.5281/zenodo.804967}

\bibitem[{Barkat {et~al.}(1967)Barkat, Rakavy, \& Sack}]{Barkat1967}
Barkat, Z., Rakavy, G., \& Sack, N. 1967, Phys. Rev. Lett., 18, 379, \dodoi{10.1103/PhysRevLett.18.379}

\bibitem[{{Barmentloo} {et~al.}(2024){Barmentloo}, {Jerkstrand}, {Iwamoto}, {Hachisu}, {Nomoto}, {Sollerman}, \& {Woosley}}]{Barmentloo2024}
{Barmentloo}, S., {Jerkstrand}, A., {Iwamoto}, K., {et~al.} 2024, \mnras, 533, 1251, \dodoi{10.1093/mnras/stae1811}

\bibitem[{{Becker}(2015)}]{Becker2015a}
{Becker}, A. 2015, {HOTPANTS: High Order Transform of PSF ANd Template Subtraction}, Astrophysics Source Code Library, record ascl:1504.004

\bibitem[{{Bellm} {et~al.}(2019){Bellm}, {Kulkarni}, {Barlow}, {Feindt}, {Graham}, {Goobar}, {Kupfer}, {Ngeow}, {Nugent}, {Ofek}, {Prince}, {Riddle}, {Walters}, \& {Ye}}]{Bellm2019b}
{Bellm}, E.~C., {Kulkarni}, S.~R., {Barlow}, T., {et~al.} 2019, \pasp, 131, 068003, \dodoi{10.1088/1538-3873/ab0c2a}

\bibitem[{{Ben-Ami} {et~al.}(2012){Ben-Ami}, {Konidaris}, {Quimby}, {Davis}, {Ngeow}, {Ritter}, \& {Rudy}}]{Ben-Ami12}
{Ben-Ami}, S., {Konidaris}, N., {Quimby}, R., {et~al.} 2012, in \procspie, Vol. 8446, Ground-based and Airborne Instrumentation for Astronomy IV, 844686, \dodoi{10.1117/12.926317}

\bibitem[{{Bersten} {et~al.}(2013){Bersten}, {Tanaka}, {Tominaga}, {Benvenuto}, \& {Nomoto}}]{Bersten2013}
{Bersten}, M.~C., {Tanaka}, M., {Tominaga}, N., {Benvenuto}, O.~G., \& {Nomoto}, K. 2013, \apj, 767, 143, \dodoi{10.1088/0004-637X/767/2/143}

\bibitem[{{Blagorodnova} {et~al.}(2018){Blagorodnova}, {Neill}, {Walters}, {Kulkarni}, {Fremling}, {Ben-Ami}, {Dekany}, {Fucik}, {Konidaris}, {Nash}, {Ngeow}, {Ofek}, {O' Sullivan}, {Quimby}, {Ritter}, \& {Vyhmeister}}]{sedm2018}
{Blagorodnova}, N., {Neill}, J.~D., {Walters}, R., {et~al.} 2018, \pasp, 130, 035003, \dodoi{10.1088/1538-3873/aaa53f}

\bibitem[{{Blondin} \& {Tonry}(2007)}]{snid2007}
{Blondin}, S., \& {Tonry}, J.~L. 2007, \apj, 666, 1024, \dodoi{10.1086/520494}

\bibitem[{{Breeveld} {et~al.}(2011){Breeveld}, {Landsman}, {Holland}, {Roming}, {Kuin}, \& {Page}}]{Breeveld2011a}
{Breeveld}, A.~A., {Landsman}, W., {Holland}, S.~T., {et~al.} 2011, in American Institute of Physics Conference Series, Vol. 1358, American Institute of Physics Conference Series, ed. J.~E. {McEnery}, J.~L. {Racusin}, \& N.~{Gehrels}, 373--376, \dodoi{10.1063/1.3621807}

\bibitem[{{Buzzoni} {et~al.}(1984){Buzzoni}, {Delabre}, {Dekker}, {Dodorico}, {Enard}, {Focardi}, {Gustafsson}, {Nees}, {Paureau}, \& {Reiss}}]{efosc2}
{Buzzoni}, B., {Delabre}, B., {Dekker}, H., {et~al.} 1984, The Messenger, 38, 9

\bibitem[{{Cartier} {et~al.}(2024){Cartier}, {Contreras}, {Stritzinger}, {Hamuy}, {Ruiz-Lapuente}, {Prieto}, {Anderson}, {Cikota}, \& {Gerlach}}]{Cartier2024arXiv241021381C}
{Cartier}, R., {Contreras}, C., {Stritzinger}, M., {et~al.} 2024, arXiv e-prints, arXiv:2410.21381, \dodoi{10.48550/arXiv.2410.21381}

\bibitem[{{Cenko} {et~al.}(2006){Cenko}, {Fox}, {Moon}, {Harrison}, {Kulkarni}, {Henning}, {Guzman}, {Bonati}, {Smith}, {Thicksten}, {Doyle}, {Petrie}, {Gal-Yam}, {Soderberg}, {Anagnostou}, \& {Laity}}]{Cenko2006}
{Cenko}, S.~B., {Fox}, D.~B., {Moon}, D.-S., {et~al.} 2006, \pasp, 118, 1396, \dodoi{10.1086/508366}

\bibitem[{{Chen} {et~al.}(2024){Chen}, {Gal-Yam}, {Sollerman}, {Schulze}, {Post}, {Liu}, {Ofek}, {Das}, {Fremling}, {Horesh}, {Katz}, {Kushnir}, {Kasliwal}, {Kulkarni}, {Liu}, {Liu}, {Miller}, {Rose}, {Waxman}, {Yang}, {Yao}, {Zackay}, {Bellm}, {Dekany}, {Drake}, {Fang}, {Fynbo}, {Groom}, {Helou}, {Irani}, {Jegou du Laz}, {Liu}, {Mazzali}, {Neill}, {Qin}, {Riddle}, {Sharon}, {Strotjohann}, {Wold}, \& {Yan}}]{Chen2024Natur.625..253C}
{Chen}, P., {Gal-Yam}, A., {Sollerman}, J., {et~al.} 2024, \nat, 625, 253, \dodoi{10.1038/s41586-023-06787-x}

\bibitem[{{Chen}(2019)}]{epessto}
{Chen}, T.-W. 2019, in The Extragalactic Explosive Universe: the New Era of Transient Surveys and Data-Driven Discovery, 15, \dodoi{10.5281/zenodo.3478022}

\bibitem[{{Chen} {et~al.}(2023{\natexlab{a}}){Chen}, {Yan}, {Kangas}, {Lunnan}, {Schulze}, {Sollerman}, {Perley}, {Chen}, {Taggart}, {Hinds}, {Gal-Yam}, {Wang}, {Andreoni}, {Bellm}, {Bloom}, {Burdge}, {Burgos}, {Cook}, {Dahiwale}, {De}, {Dekany}, {Dugas}, {Frederik}, {Fremling}, {Graham}, {Hankins}, {Ho}, {Jencson}, {Karambelkar}, {Kasliwal}, {Kulkarni}, {Laher}, {Rusholme}, {Sharma}, {Taddia}, {Tartaglia}, {Thomas}, {Tzanidakis}, {Van Roestel}, {Walter}, {Yang}, {Yao}, \& {Yaron}}]{ChenYan2023}
{Chen}, Z.~H., {Yan}, L., {Kangas}, T., {et~al.} 2023{\natexlab{a}}, \apj, 943, 41, \dodoi{10.3847/1538-4357/aca161}

\bibitem[{{Chen} {et~al.}(2023{\natexlab{b}}){Chen}, {Yan}, {Kangas}, {Lunnan}, {Sollerman}, {Schulze}, {Perley}, {Chen}, {Taggart}, {Hinds}, {Gal-Yam}, {Wang}, {De}, {Bellm}, {Bloom}, {Dekany}, {Graham}, {Kasliwal}, {Kulkarni}, {Laher}, {Neill}, \& {Rusholme}}]{ChenSLSNZTFa2023ApJ...943...42C}
---. 2023{\natexlab{b}}, \apj, 943, 42, \dodoi{10.3847/1538-4357/aca162}

\bibitem[{{Chiba} \& {Moriya}(2024)}]{Chiba2024ApJ...973...14C}
{Chiba}, R., \& {Moriya}, T.~J. 2024, \apj, 973, 14, \dodoi{10.3847/1538-4357/ad6c37}

\bibitem[{{Chu} {et~al.}(2021){Chu}, {Dahiwale}, \& {Fremling}}]{ChuTNS2021}
{Chu}, M., {Dahiwale}, A., \& {Fremling}, C. 2021, Transient Name Server Classification Report, 2021-3171, 1

\bibitem[{{Chugai} \& {Utrobin}(2022)}]{Chugaimagnetar2022MNRAS.512L..71C}
{Chugai}, N.~N., \& {Utrobin}, V.~P. 2022, \mnras, 512, L71, \dodoi{10.1093/mnrasl/slab131}

\bibitem[{{Coughlin} {et~al.}(2023){Coughlin}, {Bloom}, {Nir}, {Antier}, {du Laz}, {van der Walt}, {Crellin-Quick}, {Culino}, {Duev}, {Goldstein}, {Healy}, {Karambelkar}, {Lilleboe}, {Shin}, {Singer}, {Ahumada}, {Anand}, {Bellm}, {Dekany}, {Graham}, {Kasliwal}, {Kostadinova}, {Kiendrebeogo}, {Kulkarni}, {Jenkins}, {LeBaron}, {Mahabal}, {Neill}, {Parazin}, {Peloton}, {Perley}, {Riddle}, {Rusholme}, {van Santen}, {Sollerman}, {Stein}, {Turpin}, {Wold}, {Amat}, {Bonnefon}, {Bonnefoy}, {Flament}, {Kerkow}, {Kishore}, {Jani}, {Mahanty}, {Liu}, {Llinares}, {Makarison}, {Olli{\'e}ric}, {Perez}, {Pont}, \& {Sharma}}]{Coughlin2023}
{Coughlin}, M.~W., {Bloom}, J.~S., {Nir}, G., {et~al.} 2023, \apjs, 267, 31, \dodoi{10.3847/1538-4365/acdee1}

\bibitem[{{Crawford} {et~al.}(2025){Crawford}, {Pritchard}, {Modjaz}, {Pellegrino}, {Kumar}, \& {Baer-Way}}]{Crawford2025arXiv250303735C}
{Crawford}, A., {Pritchard}, T.~A., {Modjaz}, M., {et~al.} 2025, arXiv e-prints, arXiv:2503.03735, \dodoi{10.48550/arXiv.2503.03735}

\bibitem[{{Das} {et~al.}(2024){Das}, {Kasliwal}, {Sollerman}, {Fremling}, {Irani}, {Leung}, {Yang}, {Wu}, {Fuller}, {Anand}, {Andreoni}, {Barbarino}, {Brink}, {De}, {Dugas}, {Groom}, {Helou}, {Hinds}, {Ho}, {Karambelkar}, {Kulkarni}, {Perley}, {Purdum}, {Regnault}, {Schulze}, {Sharma}, {Sit}, {Sravan}, {Srinivasaragavan}, {Stein}, {Taggart}, {Tartaglia}, {Tzanidakis}, {Wold}, {Yan}, {Yao}, \& {Zolkower}}]{Das2024}
{Das}, K.~K., {Kasliwal}, M.~M., {Sollerman}, J., {et~al.} 2024, \apj, 972, 91, \dodoi{10.3847/1538-4357/ad595f}

\bibitem[{{Dekany} {et~al.}(2020){Dekany}, {Smith}, {Riddle}, {Feeney}, {Porter}, {Hale}, {Zolkower}, {Belicki}, {Kaye}, {Henning}, {Walters}, {Cromer}, {Delacroix}, {Rodriguez}, {Reiley}, {Mao}, {Hover}, {Murphy}, {Burruss}, {Baker}, {Kowalski}, {Reif}, {Mueller}, {Bellm}, {Graham}, \& {Kulkarni}}]{Dekany20}
{Dekany}, R., {Smith}, R.~M., {Riddle}, R., {et~al.} 2020, \pasp, 132, 038001, \dodoi{10.1088/1538-3873/ab4ca2}

\bibitem[{{Dey} {et~al.}(2019){Dey}, {Schlegel}, {Lang}, {Blum}, {Burleigh}, {Fan}, {Findlay}, {Finkbeiner}, {Herrera}, {Juneau}, {Landriau}, {Levi}, {McGreer}, {Meisner}, {Myers}, {Moustakas}, {Nugent}, {Patej}, {Schlafly}, {Walker}, {Valdes}, {Weaver}, {Y{\`e}che}, {Zou}, {Zhou}, {Abareshi}, {Abbott}, {Abolfathi}, {Aguilera}, {Alam}, {Allen}, {Alvarez}, {Annis}, {Ansarinejad}, {Aubert}, {Beechert}, {Bell}, {BenZvi}, {Beutler}, {Bielby}, {Bolton}, {Brice{\~n}o}, {Buckley-Geer}, {Butler}, {Calamida}, {Carlberg}, {Carter}, {Casas}, {Castander}, {Choi}, {Comparat}, {Cukanovaite}, {Delubac}, {DeVries}, {Dey}, {Dhungana}, {Dickinson}, {Ding}, {Donaldson}, {Duan}, {Duckworth}, {Eftekharzadeh}, {Eisenstein}, {Etourneau}, {Fagrelius}, {Farihi}, {Fitzpatrick}, {Font-Ribera}, {Fulmer}, {G{\"a}nsicke}, {Gaztanaga}, {George}, {Gerdes}, {Gontcho}, {Gorgoni}, {Green}, {Guy}, {Harmer}, {Hernandez}, {Honscheid}, {Huang}, {James}, {Jannuzi}, {Jiang}, {Joyce}, {Karcher}, {Karkar}, {Kehoe}, {Kneib}, {Kueter-Young}, {Lan},
  {Lauer}, {Le Guillou}, {Le Van Suu}, {Lee}, {Lesser}, {Perreault Levasseur}, {Li}, {Mann}, {Marshall}, {Mart{\'\i}nez-V{\'a}zquez}, {Martini}, {du Mas des Bourboux}, {McManus}, {Meier}, {M{\'e}nard}, {Metcalfe}, {Mu{\~n}oz-Guti{\'e}rrez}, {Najita}, {Napier}, {Narayan}, {Newman}, {Nie}, {Nord}, {Norman}, {Olsen}, {Paat}, {Palanque-Delabrouille}, {Peng}, {Poppett}, {Poremba}, {Prakash}, {Rabinowitz}, {Raichoor}, {Rezaie}, {Robertson}, {Roe}, {Ross}, {Ross}, {Rudnick}, {Safonova}, {Saha}, {S{\'a}nchez}, {Savary}, {Schweiker}, {Scott}, {Seo}, {Shan}, {Silva}, {Slepian}, {Soto}, {Sprayberry}, {Staten}, {Stillman}, {Stupak}, {Summers}, {Sien Tie}, {Tirado}, {Vargas-Maga{\~n}a}, {Vivas}, {Wechsler}, {Williams}, {Yang}, {Yang}, {Yapici}, {Zaritsky}, {Zenteno}, {Zhang}, {Zhang}, {Zhou}, \& {Zhou}}]{Dey2019a}
{Dey}, A., {Schlegel}, D.~J., {Lang}, D., {et~al.} 2019, \aj, 157, 168, \dodoi{10.3847/1538-3881/ab089d}

\bibitem[{{Dong} {et~al.}(2023){Dong}, {Liu}, {Gao}, \& {Yang}}]{Dong2023ApJ...951...61D}
{Dong}, X.-F., {Liu}, L.-D., {Gao}, H., \& {Yang}, S. 2023, \apj, 951, 61, \dodoi{10.3847/1538-4357/acd848}

\bibitem[{{Fitzpatrick}(1999)}]{fitzp1999}
{Fitzpatrick}, E.~L. 1999, \pasp, 111, 63, \dodoi{10.1086/316293}

\bibitem[{{Folatelli} {et~al.}(2006){Folatelli}, {Contreras}, {Phillips}, {Woosley}, {Blinnikov}, {Morrell}, {Suntzeff}, {Lee}, {Hamuy}, {Gonz{\'a}lez}, {Krzeminski}, {Roth}, {Li}, {Filippenko}, {Foley}, {Freedman}, {Madore}, {Persson}, {Murphy}, {Boissier}, {Galaz}, {Gonz{\'a}lez}, {McCarthy}, {McWilliam}, \& {Pych}}]{Folatelli2006}
{Folatelli}, G., {Contreras}, C., {Phillips}, M.~M., {et~al.} 2006, \apj, 641, 1039, \dodoi{10.1086/500531}

\bibitem[{{Fremling}(2021)}]{uvy2021TNS}
{Fremling}, C. 2021, Transient Name Server Discovery Report, 2021-2682, 1

\bibitem[{{Fremling}(2022)}]{yos2022TNS}
---. 2022, Transient Name Server Discovery Report, 2022-932, 1

\bibitem[{{Fremling} {et~al.}(2016){Fremling}, {Sollerman}, {Taddia}, {Ergon}, {Fraser}, {Karamehmetoglu}, {Valenti}, {Jerkstrand}, {Arcavi}, {Bufano}, {Elias Rosa}, {Filippenko}, {Fox}, {Gal-Yam}, {Howell}, {Kotak}, {Mazzali}, {Milisavljevic}, {Nugent}, {Nyholm}, {Pian}, \& {Smartt}}]{Fremling2016}
{Fremling}, C., {Sollerman}, J., {Taddia}, F., {et~al.} 2016, \aap, 593, A68, \dodoi{10.1051/0004-6361/201628275}

\bibitem[{{Fremling} {et~al.}(2020){Fremling}, {Miller}, {Sharma}, {Dugas}, {Perley}, {Taggart}, {Sollerman}, {Goobar}, {Graham}, {Neill}, {Nordin}, {Rigault}, {Walters}, {Andreoni}, {Bagdasaryan}, {Belicki}, {Cannella}, {Bellm}, {Cenko}, {De}, {Dekany}, {Frederick}, {Golkhou}, {Graham}, {Helou}, {Ho}, {Kasliwal}, {Kupfer}, {Laher}, {Mahabal}, {Masci}, {Riddle}, {Rusholme}, {Schulze}, {Shupe}, {Smith}, {van Velzen}, {Yan}, {Yao}, {Zhuang}, \& {Kulkarni}}]{Fremling2020}
{Fremling}, C., {Miller}, A.~A., {Sharma}, Y., {et~al.} 2020, \apj, 895, 32, \dodoi{10.3847/1538-4357/ab8943}

\bibitem[{{Galama} {et~al.}(1998){Galama}, {Vreeswijk}, {van Paradijs}, {Kouveliotou}, {Augusteijn}, {B{\"o}hnhardt}, {Brewer}, {Doublier}, {Gonzalez}, {Leibundgut}, {Lidman}, {Hainaut}, {Patat}, {Heise}, {in't Zand}, {Hurley}, {Groot}, {Strom}, {Mazzali}, {Iwamoto}, {Nomoto}, {Umeda}, {Nakamura}, {Young}, {Suzuki}, {Shigeyama}, {Koshut}, {Kippen}, {Robinson}, {de Wildt}, {Wijers}, {Tanvir}, {Greiner}, {Pian}, {Palazzi}, {Frontera}, {Masetti}, {Nicastro}, {Feroci}, {Costa}, {Piro}, {Peterson}, {Tinney}, {Boyle}, {Cannon}, {Stathakis}, {Sadler}, {Begam}, \& {Ianna}}]{Galama1998}
{Galama}, T.~J., {Vreeswijk}, P.~M., {van Paradijs}, J., {et~al.} 1998, \nat, 395, 670, \dodoi{10.1038/27150}

\bibitem[{{Gehrels} {et~al.}(2004){Gehrels}, {Chincarini}, {Giommi}, {Mason}, {Nousek}, {Wells}, {White}, {Barthelmy}, {Burrows}, {Cominsky}, {Hurley}, {Marshall}, {M{\'e}sz{\'a}ros}, {Roming}, {Angelini}, {Barbier}, {Belloni}, {Campana}, {Caraveo}, {Chester}, {Citterio}, {Cline}, {Cropper}, {Cummings}, {Dean}, {Feigelson}, {Fenimore}, {Frail}, {Fruchter}, {Garmire}, {Gendreau}, {Ghisellini}, {Greiner}, {Hill}, {Hunsberger}, {Krimm}, {Kulkarni}, {Kumar}, {Lebrun}, {Lloyd-Ronning}, {Markwardt}, {Mattson}, {Mushotzky}, {Norris}, {Osborne}, {Paczynski}, {Palmer}, {Park}, {Parsons}, {Paul}, {Rees}, {Reynolds}, {Rhoads}, {Sasseen}, {Schaefer}, {Short}, {Smale}, {Smith}, {Stella}, {Tagliaferri}, {Takahashi}, {Tashiro}, {Townsley}, {Tueller}, {Turner}, {Vietri}, {Voges}, {Ward}, {Willingale}, {Zerbi}, \& {Zhang}}]{Gehrels2004a}
{Gehrels}, N., {Chincarini}, G., {Giommi}, P., {et~al.} 2004, \apj, 611, 1005, \dodoi{10.1086/422091}

\bibitem[{{Gomez} {et~al.}(2021){Gomez}, {Berger}, {Hosseinzadeh}, {Blanchard}, {Nicholl}, \& {Villar}}]{Gomez2021}
{Gomez}, S., {Berger}, E., {Hosseinzadeh}, G., {et~al.} 2021, \apj, 913, 143, \dodoi{10.3847/1538-4357/abf5e3}

\bibitem[{{Gomez} {et~al.}(2022){Gomez}, {Berger}, {Nicholl}, {Blanchard}, \& {Hosseinzadeh}}]{Gomez2022}
{Gomez}, S., {Berger}, E., {Nicholl}, M., {Blanchard}, P.~K., \& {Hosseinzadeh}, G. 2022, \apj, 941, 107, \dodoi{10.3847/1538-4357/ac9842}

\bibitem[{{Graham} {et~al.}(2019){Graham}, {Kulkarni}, {Bellm}, {Adams}, {Barbarino}, {Blagorodnova}, {Bodewits}, {Bolin}, {Brady}, {Cenko}, {Chang}, {Coughlin}, {De}, {Eadie}, {Farnham}, {Feindt}, {Franckowiak}, {Fremling}, {Gezari}, {Ghosh}, {Goldstein}, {Golkhou}, {Goobar}, {Ho}, {Huppenkothen}, {Ivezi{\'c}}, {Jones}, {Juric}, {Kaplan}, {Kasliwal}, {Kelley}, {Kupfer}, {Lee}, {Lin}, {Lunnan}, {Mahabal}, {Miller}, {Ngeow}, {Nugent}, {Ofek}, {Prince}, {Rauch}, {van Roestel}, {Schulze}, {Singer}, {Sollerman}, {Taddia}, {Yan}, {Ye}, {Yu}, {Barlow}, {Bauer}, {Beck}, {Belicki}, {Biswas}, {Brinnel}, {Brooke}, {Bue}, {Bulla}, {Burruss}, {Connolly}, {Cromer}, {Cunningham}, {Dekany}, {Delacroix}, {Desai}, {and}, {Feeney}, {Flynn}, {Frederick}, {Gal-Yam}, {Giomi}, {Groom}, {Hacopians}, {Hale}, {Helou}, {Henning}, {Hover}, {Hillenbrand}, {Howell}, {Hung}, {Imel}, {Ip}, {Jackson}, {Kaspi}, {Kaye}, {Kowalski}, {Kramer}, {Kuhn}, {Landry}, {Laher}, {Mao}, {Masci}, {Monkewitz}, {Murphy}, {Nordin}, {Patterson}, {Penprase},
  {Porter}, {Rebbapragada}, {Reiley}, {Riddle}, {Rigault}, {Rodriguez}, {Rusholme}, {van Santen}, {Shupe}, {Smith}, {Soumagnac}, {Stein}, {Surace}, {Szkody}, {Terek}, {Van Sistine}, {van Velzen}, {Vestrand}, {Walters}, {Ward}, {Zhang}, \& {Zolkower}}]{graham2019}
{Graham}, M.~J., {Kulkarni}, S.~R., {Bellm}, E.~C., {et~al.} 2019, \pasp, 131, 078001, \dodoi{10.1088/1538-3873/ab006c}

\bibitem[{{Guti{\'e}rrez} {et~al.}(2021){Guti{\'e}rrez}, {Bersten}, {Orellana}, {Pastorello}, {Ertini}, {Folatelli}, {Pignata}, {Anderson}, {Smartt}, {Sullivan}, {Pursiainen}, {Inserra}, {Elias-Rosa}, {Fraser}, {Kankare}, {Moran}, {Reguitti}, {Reynolds}, {Stritzinger}, {Burke}, {Frohmaier}, {Galbany}, {Hiramatsu}, {Howell}, {Kuncarayakti}, {Mattila}, {M{\"u}ller-Bravo}, {Pellegrino}, \& {Smith}}]{Gutierrez2021}
{Guti{\'e}rrez}, C.~P., {Bersten}, M.~C., {Orellana}, M., {et~al.} 2021, \mnras, 504, 4907, \dodoi{10.1093/mnras/stab1009}

\bibitem[{Hunter(2007)}]{matplotlib}
Hunter, J.~D. 2007, Computing in Science \& Engineering, 9, 90, \dodoi{10.1109/MCSE.2007.55}

\bibitem[{{Ivezi{\'c}} {et~al.}(2019){Ivezi{\'c}}, {Kahn}, {Tyson}, {Abel}, {Acosta}, {Allsman}, {Alonso}, {AlSayyad}, {Anderson}, {Andrew}, \& et~al.}]{lsst}
{Ivezi{\'c}}, {\v Z}., {Kahn}, S.~M., {Tyson}, J.~A., {et~al.} 2019, \apj, 873, 111, \dodoi{10.3847/1538-4357/ab042c}

\bibitem[{{Jin} {et~al.}(2021){Jin}, {Yoon}, \& {Blinnikov}}]{Jin2021ApJ...910...68J}
{Jin}, H., {Yoon}, S.-C., \& {Blinnikov}, S. 2021, \apj, 910, 68, \dodoi{10.3847/1538-4357/abe0b1}

\bibitem[{{Kangas} {et~al.}(2024){Kangas}, {Kuncarayakti}, {Nagao}, {Kotak}, {Kankare}, {Fraser}, {Stevance}, {Mattila}, {Maeda}, {Stritzinger}, {Lundqvist}, {Elias-Rosa}, {Ferrari}, {Folatelli}, {Frohmaier}, {Galbany}, {Kawabata}, {Koutsiona}, {M{\"u}ller-Bravo}, {Piscarreta}, {Pursiainen}, {Singh}, {Taguchi}, {Teja}, {Valerin}, {Pastorello}, {Benetti}, {Cai}, {Charalampopoulos}, {Guti{\'e}rrez}, {Kravtsov}, \& {Reguitti}}]{Kangas2024}
{Kangas}, T., {Kuncarayakti}, H., {Nagao}, T., {et~al.} 2024, \aap, 689, A182, \dodoi{10.1051/0004-6361/202449420}

\bibitem[{{Karamehmetoglu} {et~al.}(2023){Karamehmetoglu}, {Sollerman}, {Taddia}, {Barbarino}, {Feindt}, {Fremling}, {Gal-Yam}, {Kasliwal}, {Petrushevska}, {Schulze}, {Stritzinger}, \& {Zapartas}}]{Karamehmetoglu2023A&A...678A..87K}
{Karamehmetoglu}, E., {Sollerman}, J., {Taddia}, F., {et~al.} 2023, \aap, 678, A87, \dodoi{10.1051/0004-6361/202245231}

\bibitem[{Kasliwal {et~al.}(2019)Kasliwal, Cannella, Bagdasaryan, Hung, Feindt, Singer, Coughlin, Fremling, Walters, Duev, Itoh, \& Quimby}]{Kasliwal_2019}
Kasliwal, M.~M., Cannella, C., Bagdasaryan, A., {et~al.} 2019, \pasp, 131, 038003, \dodoi{10.1088/1538-3873/aafbc2}

\bibitem[{{Khatami} \& {Kasen}(2019)}]{KhatamiKasen2019ApJ}
{Khatami}, D.~K., \& {Kasen}, D.~N. 2019, \apj, 878, 56, \dodoi{10.3847/1538-4357/ab1f09}

\bibitem[{{Khatami} \& {Kasen}(2024)}]{KhatamiKasen2024ApJ...972..140K}
---. 2024, \apj, 972, 140, \dodoi{10.3847/1538-4357/ad60c0}

\bibitem[{{Kim} {et~al.}(2022){Kim}, {Rigault}, {Neill}, {Briday}, {Copin}, {Lezmy}, {Nicolas}, {Riddle}, {Sharma}, {Smith}, {Sollerman}, \& {Walters}}]{Kim2022}
{Kim}, Y.~L., {Rigault}, M., {Neill}, J.~D., {et~al.} 2022, \pasp, 134, 024505, \dodoi{10.1088/1538-3873/ac50a0}

\bibitem[{{Kuncarayakti} {et~al.}(2015){Kuncarayakti}, {Maeda}, {Bersten}, {Folatelli}, {Morrell}, {Hsiao}, {Gonz{\'a}lez-Gait{\'a}n}, {Anderson}, {Hamuy}, {de Jaeger}, {Guti{\'e}rrez}, \& {Kawabata}}]{Kuncarayakti2015}
{Kuncarayakti}, H., {Maeda}, K., {Bersten}, M.~C., {et~al.} 2015, \aap, 579, A95, \dodoi{10.1051/0004-6361/201425604}

\bibitem[{{Kuncarayakti} {et~al.}(2023){Kuncarayakti}, {Sollerman}, {Izzo}, {Maeda}, {Yang}, {Schulze}, {Angus}, {Aubert}, {Auchettl}, {Della Valle}, {Dessart}, {Hinds}, {Kankare}, {Kawabata}, {Lundqvist}, {Nakaoka}, {Perley}, {Raimundo}, {Strotjohann}, {Taguchi}, {Cai}, {Charalampopoulos}, {Fang}, {Fraser}, {Guti{\'e}rrez}, {Imazawa}, {Kangas}, {Kawabata}, {Kotak}, {Kravtsov}, {Matilainen}, {Mattila}, {Moran}, {Murata}, {Salmaso}, {Anderson}, {Ashall}, {Bellm}, {Benetti}, {Chambers}, {Chen}, {Coughlin}, {De Colle}, {Fremling}, {Galbany}, {Gal-Yam}, {Gromadzki}, {Groom}, {Hajela}, {Inserra}, {Kasliwal}, {Mahabal}, {Martin-Carrillo}, {Moore}, {M{\"u}ller-Bravo}, {Nicholl}, {Ragosta}, {Riddle}, {Sharma}, {Srivastav}, {Stritzinger}, {Wold}, \& {Young}}]{Kuncarayakti2023}
{Kuncarayakti}, H., {Sollerman}, J., {Izzo}, L., {et~al.} 2023, \aap, 678, A209, \dodoi{10.1051/0004-6361/202346526}

\bibitem[{{Leung} {et~al.}(2019){Leung}, {Nomoto}, \& {Blinnikov}}]{2019ApJ...887...72L}
{Leung}, S.-C., {Nomoto}, K., \& {Blinnikov}, S. 2019, \apj, 887, 72, \dodoi{10.3847/1538-4357/ab4fe5}

\bibitem[{{Lunnan} {et~al.}(2021){Lunnan}, {Yan}, {Perley}, {Chen}, {Schulze}, {Kangas}, {Sollerman}, \& {Gal-Yam}}]{LunnanTNS2021}
{Lunnan}, R., {Yan}, L., {Perley}, D.~A., {et~al.} 2021, Transient Name Server AstroNote, 218, 1

\bibitem[{{Lunnan} {et~al.}(2016){Lunnan}, {Chornock}, {Berger}, {Milisavljevic}, {Jones}, {Rest}, {Fong}, {Fransson}, {Margutti}, {Drout}, {Blanchard}, {Challis}, {Cowperthwaite}, {Foley}, {Kirshner}, {Morrell}, {Riess}, {Roth}, {Scolnic}, {Smartt}, {Smith}, {Villar}, {Chambers}, {Draper}, {Huber}, {Kaiser}, {Kudritzki}, {Magnier}, {Metcalfe}, \& {Waters}}]{Lunnan2016}
{Lunnan}, R., {Chornock}, R., {Berger}, E., {et~al.} 2016, \apj, 831, 144, \dodoi{10.3847/0004-637X/831/2/144}

\bibitem[{{Maeda} {et~al.}(2007){Maeda}, {Tanaka}, {Nomoto}, {Tominaga}, {Kawabata}, {Mazzali}, {Umeda}, {Suzuki}, \& {Hattori}}]{Maeda2007}
{Maeda}, K., {Tanaka}, M., {Nomoto}, K., {et~al.} 2007, \apj, 666, 1069, \dodoi{10.1086/520054}

\bibitem[{{Masci} {et~al.}(2019){Masci}, {Laher}, {Rusholme}, {Shupe}, {Groom}, {Surace}, {Jackson}, {Monkewitz}, {Beck}, {Flynn}, {Terek}, {Landry}, {Hacopians}, {Desai}, {Howell}, {Brooke}, {Imel}, {Wachter}, {Ye}, {Lin}, {Cenko}, {Cunningham}, {Rebbapragada}, {Bue}, {Miller}, {Mahabal}, {Bellm}, {Patterson}, {Juri{\'c}}, {Golkhou}, {Ofek}, {Walters}, {Graham}, {Kasliwal}, {Dekany}, {Kupfer}, {Burdge}, {Cannella}, {Barlow}, {Van Sistine}, {Giomi}, {Fremling}, {Blagorodnova}, {Levitan}, {Riddle}, {Smith}, {Helou}, {Prince}, \& {Kulkarni}}]{Masci2019}
{Masci}, F.~J., {Laher}, R.~R., {Rusholme}, B., {et~al.} 2019, \pasp, 131, 018003, \dodoi{10.1088/1538-3873/aae8ac}

\bibitem[{{Masci} {et~al.}(2023){Masci}, {Laher}, {Rusholme}, {Shupe}, {Paladini}, {Groom}, {Wold}, {Miller}, \& {Drake}}]{Masci2023}
---. 2023, arXiv e-prints, arXiv:2305.16279, \dodoi{10.48550/arXiv.2305.16279}

\bibitem[{{Milisavljevic} {et~al.}(2010){Milisavljevic}, {Fesen}, {Gerardy}, {Kirshner}, \& {Challis}}]{Milisavljevic2010ApJ}
{Milisavljevic}, D., {Fesen}, R.~A., {Gerardy}, C.~L., {Kirshner}, R.~P., \& {Challis}, P. 2010, \apj, 709, 1343, \dodoi{10.1088/0004-637X/709/2/1343}

\bibitem[{{Morales-Garoffolo} {et~al.}(2015){Morales-Garoffolo}, {Elias-Rosa}, {Bersten}, {Jerkstrand}, {Taubenberger}, {Benetti}, {Cappellaro}, {Kotak}, {Pastorello}, {Bufano}, {Dom{\'\i}nguez}, {Ergon}, {Fraser}, {Gao}, {Garc{\'\i}a}, {Howell}, {Isern}, {Smartt}, {Tomasella}, \& {Valenti}}]{Morales2015MNRAS.454...95M}
{Morales-Garoffolo}, A., {Elias-Rosa}, N., {Bersten}, M., {et~al.} 2015, \mnras, 454, 95, \dodoi{10.1093/mnras/stv1972}

\bibitem[{{Moriya} {et~al.}(2022){Moriya}, {Murase}, {Kashiyama}, \& {Blinnikov}}]{MoriyaMagnetar2022MNRAS.513.6210M}
{Moriya}, T.~J., {Murase}, K., {Kashiyama}, K., \& {Blinnikov}, S.~I. 2022, \mnras, 513, 6210, \dodoi{10.1093/mnras/stac1352}

\bibitem[{{Nakar} \& {Piro}(2014)}]{PiroNakar2014ApJ...788..193N}
{Nakar}, E., \& {Piro}, A.~L. 2014, \apj, 788, 193, \dodoi{10.1088/0004-637X/788/2/193}

\bibitem[{{Nicholl}(2018)}]{Nicholl2018RNAAS...2..230N}
{Nicholl}, M. 2018, Research Notes of the American Astronomical Society, 2, 230, \dodoi{10.3847/2515-5172/aaf799}

\bibitem[{{Oke} \& {Gunn}(1982)}]{Oke1982}
{Oke}, J.~B., \& {Gunn}, J.~E. 1982, \pasp, 94, 586, \dodoi{10.1086/131027}

\bibitem[{{Oke} {et~al.}(1995){Oke}, {Cohen}, {Carr}, {Cromer}, {Dingizian}, {Harris}, {Labrecque}, {Lucinio}, {Schaal}, {Epps}, \& {Miller}}]{Oke1995}
{Oke}, J.~B., {Cohen}, J.~G., {Carr}, M., {et~al.} 1995, \pasp, 107, 375, \dodoi{10.1086/133562}

\bibitem[{{Orellana} \& {Bersten}(2022)}]{Orellana2022A&A...667A..92O}
{Orellana}, M., \& {Bersten}, M.~C. 2022, \aap, 667, A92, \dodoi{10.1051/0004-6361/202244124}

\bibitem[{{Patat} {et~al.}(2001){Patat}, {Cappellaro}, {Danziger}, {Mazzali}, {Sollerman}, {Augusteijn}, {Brewer}, {Doublier}, {Gonzalez}, {Hainaut}, {Lidman}, {Leibundgut}, {Nomoto}, {Nakamura}, {Spyromilio}, {Rizzi}, {Turatto}, {Walsh}, {Galama}, {van Paradijs}, {Kouveliotou}, {Vreeswijk}, {Frontera}, {Masetti}, {Palazzi}, \& {Pian}}]{Patat2001}
{Patat}, F., {Cappellaro}, E., {Danziger}, J., {et~al.} 2001, \apj, 555, 900, \dodoi{10.1086/321526}

\bibitem[{{Pellegrino} {et~al.}(2023){Pellegrino}, {Hiramatsu}, {Arcavi}, {Howell}, {Bostroem}, {Brown}, {Burke}, {Elias-Rosa}, {Itagaki}, {Kaneda}, {McCully}, {Modjaz}, {Padilla Gonzalez}, {Pritchard}, \& {Yesmin}}]{Pellegrino2023ApJ...954...35P}
{Pellegrino}, C., {Hiramatsu}, D., {Arcavi}, I., {et~al.} 2023, \apj, 954, 35, \dodoi{10.3847/1538-4357/ace595}

\bibitem[{{Perley} {et~al.}(2022){Perley}, {Chu}, {Dahiwale}, \& {Fremling}}]{yos2022TNS_classification}
{Perley}, D., {Chu}, M., {Dahiwale}, A., \& {Fremling}, C. 2022, Transient Name Server Classification Report, 2022-1362, 1

\bibitem[{{Perley}(2019)}]{Perley2019}
{Perley}, D.~A. 2019, \pasp, 131, 084503, \dodoi{10.1088/1538-3873/ab215d}

\bibitem[{Perley {et~al.}(2020)Perley, Fremling, Sollerman, Miller, Dahiwale, Sharma, Bellm, Biswas, Brink, Bruch, \& et~al.}]{Perley2020}
Perley, D.~A., Fremling, C., Sollerman, J., {et~al.} 2020, \apj, 904, 35, \dodoi{10.3847/1538-4357/abbd98}

\bibitem[{{Piascik} {et~al.}(2014){Piascik}, {Steele}, {Bates}, {Mottram}, {Smith}, {Barnsley}, \& {Bolton}}]{SPRAT}
{Piascik}, A.~S., {Steele}, I.~A., {Bates}, S.~D., {et~al.} 2014, in \procspie, Vol. 9147, Ground-based and Airborne Instrumentation for Astronomy V, 91478H, \dodoi{10.1117/12.2055117}

\bibitem[{{Piro}(2015)}]{Piro2015ApJ...808L..51P}
{Piro}, A.~L. 2015, \apjl, 808, L51, \dodoi{10.1088/2041-8205/808/2/L51}

\bibitem[{{Piro} {et~al.}(2021){Piro}, {Haynie}, \& {Yao}}]{Piro2021ApJ...909..209P}
{Piro}, A.~L., {Haynie}, A., \& {Yao}, Y. 2021, \apj, 909, 209, \dodoi{10.3847/1538-4357/abe2b1}

\bibitem[{{Planck Collaboration} {et~al.}(2020){Planck Collaboration}, {Aghanim}, {Akrami}, {Ashdown}, {Aumont}, {Baccigalupi}, {Ballardini}, {Banday}, {Barreiro}, {Bartolo}, {Basak}, {Battye}, {Benabed}, {Bernard}, {Bersanelli}, {Bielewicz}, {Bock}, {Bond}, {Borrill}, {Bouchet}, {Boulanger}, {Bucher}, {Burigana}, {Butler}, {Calabrese}, {Cardoso}, {Carron}, {Challinor}, {Chiang}, {Chluba}, {Colombo}, {Combet}, {Contreras}, {Crill}, {Cuttaia}, {de Bernardis}, {de Zotti}, {Delabrouille}, {Delouis}, {Di Valentino}, {Diego}, {Dor{\'e}}, {Douspis}, {Ducout}, {Dupac}, {Dusini}, {Efstathiou}, {Elsner}, {En{\ss}lin}, {Eriksen}, {Fantaye}, {Farhang}, {Fergusson}, {Fernandez-Cobos}, {Finelli}, {Forastieri}, {Frailis}, {Fraisse}, {Franceschi}, {Frolov}, {Galeotta}, {Galli}, {Ganga}, {G{\'e}nova-Santos}, {Gerbino}, {Ghosh}, {Gonz{\'a}lez-Nuevo}, {G{\'o}rski}, {Gratton}, {Gruppuso}, {Gudmundsson}, {Hamann}, {Handley}, {Hansen}, {Herranz}, {Hildebrandt}, {Hivon}, {Huang}, {Jaffe}, {Jones}, {Karakci}, {Keih{\"a}nen},
  {Keskitalo}, {Kiiveri}, {Kim}, {Kisner}, {Knox}, {Krachmalnicoff}, {Kunz}, {Kurki-Suonio}, {Lagache}, {Lamarre}, {Lasenby}, {Lattanzi}, {Lawrence}, {Le Jeune}, {Lemos}, {Lesgourgues}, {Levrier}, {Lewis}, {Liguori}, {Lilje}, {Lilley}, {Lindholm}, {L{\'o}pez-Caniego}, {Lubin}, {Ma}, {Mac{\'\i}as-P{\'e}rez}, {Maggio}, {Maino}, {Mandolesi}, {Mangilli}, {Marcos-Caballero}, {Maris}, {Martin}, {Martinelli}, {Mart{\'\i}nez-Gonz{\'a}lez}, {Matarrese}, {Mauri}, {McEwen}, {Meinhold}, {Melchiorri}, {Mennella}, {Migliaccio}, {Millea}, {Mitra}, {Miville-Desch{\^e}nes}, {Molinari}, {Montier}, {Morgante}, {Moss}, {Natoli}, {N{\o}rgaard-Nielsen}, {Pagano}, {Paoletti}, {Partridge}, {Patanchon}, {Peiris}, {Perrotta}, {Pettorino}, {Piacentini}, {Polastri}, {Polenta}, {Puget}, {Rachen}, {Reinecke}, {Remazeilles}, {Renzi}, {Rocha}, {Rosset}, {Roudier}, {Rubi{\~n}o-Mart{\'\i}n}, {Ruiz-Granados}, {Salvati}, {Sandri}, {Savelainen}, {Scott}, {Shellard}, {Sirignano}, {Sirri}, {Spencer}, {Sunyaev}, {Suur-Uski}, {Tauber}, {Tavagnacco},
  {Tenti}, {Toffolatti}, {Tomasi}, {Trombetti}, {Valenziano}, {Valiviita}, {Van Tent}, {Vibert}, {Vielva}, {Villa}, {Vittorio}, {Wandelt}, {Wehus}, {White}, {White}, {Zacchei}, \& {Zonca}}]{Planck2020}
{Planck Collaboration}, {Aghanim}, N., {Akrami}, Y., {et~al.} 2020, \aap, 641, A6, \dodoi{10.1051/0004-6361/201833910}

\bibitem[{{Poidevin} {et~al.}(2021){Poidevin}, {Perez-Fournon}, {Angel}, {Shirley}, {Marques-Chaves}, {Geier}, {Shu}, {Rodney}, {Roberts-Pierel}, {Bolton}, {Chakrabarti}, {Craig}, \& {Alamiri}}]{PoidevinTNS2021}
{Poidevin}, F., {Perez-Fournon}, I., {Angel}, C.~J., {et~al.} 2021, Transient Name Server Classification Report, 2021-2800, 1

\bibitem[{{Prentice} {et~al.}(2022){Prentice}, {Maguire}, {Siebenaler}, \& {Jerkstrand}}]{2022MNRAS.514.5686P}
{Prentice}, S.~J., {Maguire}, K., {Siebenaler}, L., \& {Jerkstrand}, A. 2022, \mnras, 514, 5686, \dodoi{10.1093/mnras/stac1657}

\bibitem[{{Prentice} {et~al.}(2019){Prentice}, {Ashall}, {James}, {Short}, {Mazzali}, {Bersier}, {Crowther}, {Barbarino}, {Chen}, {Copperwheat}, {Darnley}, {Denneau}, {Elias-Rosa}, {Fraser}, {Galbany}, {Gal-Yam}, {Harmanen}, {Howell}, {Hosseinzadeh}, {Inserra}, {Kankare}, {Karamehmetoglu}, {Lamb}, {Limongi}, {Maguire}, {McCully}, {Olivares E}, {Piascik}, {Pignata}, {Reichart}, {Rest}, {Reynolds}, {Rodr{\'\i}guez}, {Saario}, {Schulze}, {Smartt}, {Smith}, {Sollerman}, {Stalder}, {Sullivan}, {Taddia}, {Valenti}, {Vergani}, {Williams}, \& {Young}}]{Prentice2019MNRAS.485.1559P}
{Prentice}, S.~J., {Ashall}, C., {James}, P.~A., {et~al.} 2019, \mnras, 485, 1559, \dodoi{10.1093/mnras/sty3399}

\bibitem[{Prochaska {et~al.}(2020)Prochaska, Hennawi, Westfall, Cooke, Wang, Hsyu, Davies, Farina, \& Pelliccia}]{pypeit:joss_pub}
Prochaska, J.~X., Hennawi, J.~F., Westfall, K.~B., {et~al.} 2020, Journal of Open Source Software, 5, 2308, \dodoi{10.21105/joss.02308}

\bibitem[{{Rakavy} {et~al.}(1967){Rakavy}, {Shaviv}, \& {Zinamon}}]{Rakavy1967}
{Rakavy}, G., {Shaviv}, G., \& {Zinamon}, Z. 1967, \apj, 150, 131, \dodoi{10.1086/149318}

\bibitem[{{Rehemtulla} {et~al.}(2024){Rehemtulla}, {Miller}, {Jegou Du Laz}, {Coughlin}, {Fremling}, {Perley}, {Qin}, {Sollerman}, {Mahabal}, {Laher}, {Riddle}, {Rusholme}, \& {Kulkarni}}]{Rehemtulla2024}
{Rehemtulla}, N., {Miller}, A.~A., {Jegou Du Laz}, T., {et~al.} 2024, \apj, 972, 7, \dodoi{10.3847/1538-4357/ad5666}

\bibitem[{{Ridley} {et~al.}(2021){Ridley}, {Gompertz}, {Nicholl}, {Galbany}, \& {Yaron}}]{RidleyTNS2021}
{Ridley}, E., {Gompertz}, B., {Nicholl}, M., {Galbany}, L., \& {Yaron}, O. 2021, Transient Name Server Classification Report, 2021-2795, 1

\bibitem[{{Rigault} {et~al.}(2019){Rigault}, {Neill}, {Blagorodnova}, {Dugas}, {Feeney}, {Walters}, {Brinnel}, {Copin}, {Fremling}, {Nordin}, \& {Sollerman}}]{pysedm2019}
{Rigault}, M., {Neill}, J.~D., {Blagorodnova}, N., {et~al.} 2019, \aap, 627, A115, \dodoi{10.1051/0004-6361/201935344}

\bibitem[{{Roberson} {et~al.}(2022){Roberson}, {Fremling}, \& {Kasliwal}}]{Roberson2022}
{Roberson}, M., {Fremling}, C., \& {Kasliwal}, M. 2022, The Journal of Open Source Software, 7, 3612, \dodoi{10.21105/joss.03612}

\bibitem[{{Rodr{\'\i}guez} {et~al.}(2024){Rodr{\'\i}guez}, {Nakar}, \& {Maoz}}]{Rodriquez2024Natur.628..733R}
{Rodr{\'\i}guez}, {\'O}., {Nakar}, E., \& {Maoz}, D. 2024, \nat, 628, 733, \dodoi{10.1038/s41586-024-07262-x}

\bibitem[{{Roming} {et~al.}(2005){Roming}, {Kennedy}, {Mason}, {Nousek}, {Ahr}, {Bingham}, {Broos}, {Carter}, {Hancock}, {Huckle}, {Hunsberger}, {Kawakami}, {Killough}, {Koch}, {McLelland}, {Smith}, {Smith}, {Soto}, {Boyd}, {Breeveld}, {Holland}, {Ivanushkina}, {Pryzby}, {Still}, \& {Stock}}]{Roming2005a}
{Roming}, P. W.~A., {Kennedy}, T.~E., {Mason}, K.~O., {et~al.} 2005, \ssr, 120, 95, \dodoi{10.1007/s11214-005-5095-4}

\bibitem[{Schlafly \& Finkbeiner(2011)}]{Schlafly_2011}
Schlafly, E.~F., \& Finkbeiner, D.~P. 2011, \apj, 737, 103, \dodoi{10.1088/0004-637x/737/2/103}

\bibitem[{{Schulze} {et~al.}(2024){Schulze}, {Fransson}, {Kozyreva}, {Chen}, {Yaron}, {Jerkstrand}, {Gal-Yam}, {Sollerman}, {Yan}, {Kangas}, {Leloudas}, {Omand}, {Smartt}, {Yang}, {Nicholl}, {Sarin}, {Yao}, {Brink}, {Sharon}, {Rossi}, {Chen}, {Chen}, {Cikota}, {De}, {Drake}, {Filippenko}, {Fremling}, {Fr{\'e}our}, {Fynbo}, {Ho}, {Inserra}, {Irani}, {Kuncarayakti}, {Lunnan}, {Mazzali}, {Ofek}, {Palazzi}, {Perley}, {Pursiainen}, {Rothberg}, {Shingles}, {Smith}, {Taggart}, {Tartaglia}, {Zheng}, {Anderson}, {Cassara}, {Christensen}, {George Djorgovski}, {Galbany}, {Gkini}, {Graham}, {Gromadzki}, {Groom}, {Hiramatsu}, {Andrew Howell}, {Kasliwal}, {McCully}, {M{\"u}ller-Bravo}, {Paiano}, {Paraskeva}, {Pessi}, {Polishook}, {Rau}, {Rigault}, \& {Rusholme}}]{Schulze2024}
{Schulze}, S., {Fransson}, C., {Kozyreva}, A., {et~al.} 2024, \aap, 683, A223, \dodoi{10.1051/0004-6361/202346855}

\bibitem[{{Sharma} {et~al.}(2024){Sharma}, {Sollerman}, {Kulkarni}, {Moriya}, {Schulze}, {Barmentloo}, {Fausnaugh}, {Gal-Yam}, {Jerkstrand}, {Ahumada}, {Bellm}, {Das}, {Drake}, {Fremling}, {Hale}, {Hall}, {Hinds}, {Jegou du Laz}, {Karambelkar}, {Kasliwal}, {Masci}, {Miller}, {Nir}, {Perley}, {Purdum}, {Qin}, {Rehemtulla}, {Rich}, {Riddle}, {Rodriguez}, {Rose}, {Somalwar}, {Wise}, {Wold}, {Yan}, \& {Yao}}]{Sharma2024}
{Sharma}, Y., {Sollerman}, J., {Kulkarni}, S.~R., {et~al.} 2024, \apj, 966, 199, \dodoi{10.3847/1538-4357/ad3758}

\bibitem[{{Shivvers} {et~al.}(2019){Shivvers}, {Filippenko}, {Silverman}, {Zheng}, {Foley}, {Chornock}, {Barth}, {Cenko}, {Clubb}, {Fox}, {Ganeshalingam}, {Graham}, {Kelly}, {Kleiser}, {Leonard}, {Li}, {Matheson}, {Mauerhan}, {Modjaz}, {Serduke}, {Shields}, {Steele}, {Swift}, {Wong}, \& {Yuk}}]{Shivvers2019}
{Shivvers}, I., {Filippenko}, A.~V., {Silverman}, J.~M., {et~al.} 2019, \mnras, 482, 1545, \dodoi{10.1093/mnras/sty2719}

\bibitem[{{Smartt} {et~al.}(2015){Smartt}, {Valenti}, {Fraser}, {Inserra}, {Young}, {Sullivan}, {Pastorello}, {Benetti}, {Gal-Yam}, {Knapic}, {Molinaro}, {Smareglia}, {Smith}, {Taubenberger}, {Yaron}, {Anderson}, {Ashall}, {Balland}, {Baltay}, {Barbarino}, {Bauer}, {Baumont}, {Bersier}, {Blagorodnova}, {Bongard}, {Botticella}, {Bufano}, {Bulla}, {Cappellaro}, {Campbell}, {Cellier-Holzem}, {Chen}, {Childress}, {Clocchiatti}, {Contreras}, {Dall'Ora}, {Danziger}, {de Jaeger}, {De Cia}, {Della Valle}, {Dennefeld}, {Elias-Rosa}, {Elman}, {Feindt}, {Fleury}, {Gall}, {Gonzalez-Gaitan}, {Galbany}, {Morales Garoffolo}, {Greggio}, {Guillou}, {Hachinger}, {Hadjiyska}, {Hage}, {Hillebrandt}, {Hodgkin}, {Hsiao}, {James}, {Jerkstrand}, {Kangas}, {Kankare}, {Kotak}, {Kromer}, {Kuncarayakti}, {Leloudas}, {Lundqvist}, {Lyman}, {Hook}, {Maguire}, {Manulis}, {Margheim}, {Mattila}, {Maund}, {Mazzali}, {McCrum}, {McKinnon}, {Moreno-Raya}, {Nicholl}, {Nugent}, {Pain}, {Pignata}, {Phillips}, {Polshaw}, {Pumo}, {Rabinowitz},
  {Reilly}, {Romero-Ca{\~n}izales}, {Scalzo}, {Schmidt}, {Schulze}, {Sim}, {Sollerman}, {Taddia}, {Tartaglia}, {Terreran}, {Tomasella}, {Turatto}, {Walker}, {Walton}, {Wyrzykowski}, {Yuan}, \& {Zampieri}}]{pessto}
{Smartt}, S.~J., {Valenti}, S., {Fraser}, M., {et~al.} 2015, \aap, 579, A40, \dodoi{10.1051/0004-6361/201425237}

\bibitem[{{Smith} {et~al.}(2020){Smith}, {Smartt}, {Young}, {Tonry}, {Denneau}, {Flewelling}, {Heinze}, {Weiland}, {Stalder}, {Rest}, {Stubbs}, {Anderson}, {Chen}, {Clark}, {Do}, {F{\"o}rster}, {Fulton}, {Gillanders}, {McBrien}, {O'Neill}, {Srivastav}, \& {Wright}}]{atlassmith2020}
{Smith}, K.~W., {Smartt}, S.~J., {Young}, D.~R., {et~al.} 2020, \pasp, 132, 085002, \dodoi{10.1088/1538-3873/ab936e}

\bibitem[{{Sollerman} {et~al.}(2022){Sollerman}, {Yang}, {Perley}, {Schulze}, {Fremling}, {Kasliwal}, {Shin}, \& {Racine}}]{Sollerman2022A&A...657A..64S}
{Sollerman}, J., {Yang}, S., {Perley}, D., {et~al.} 2022, \aap, 657, A64, \dodoi{10.1051/0004-6361/202142049}

\bibitem[{{Sollerman} {et~al.}(2002){Sollerman}, {Holland}, {Challis}, {Fransson}, {Garnavich}, {Kirshner}, {Kozma}, {Leibundgut}, {Lundqvist}, {Patat}, {Filippenko}, {Panagia}, \& {Wheeler}}]{Sollerman2002}
{Sollerman}, J., {Holland}, S.~T., {Challis}, P., {et~al.} 2002, \aap, 386, 944, \dodoi{10.1051/0004-6361:20020326}

\bibitem[{{Sollerman} {et~al.}(2020){Sollerman}, {Fransson}, {Barbarino}, {Fremling}, {Horesh}, {Kool}, {Schulze}, {Sfaradi}, {Yang}, {Bellm}, {Burruss}, {Cunningham}, {De}, {Drake}, {Golkhou}, {Green}, {Kasliwal}, {Kulkarni}, {Kupfer}, {Laher}, {Masci}, {Rodriguez}, {Rusholme}, {Williams}, {Yan}, \& {Zolkower}}]{Sollerman2020}
{Sollerman}, J., {Fransson}, C., {Barbarino}, C., {et~al.} 2020, \aap, 643, A79, \dodoi{10.1051/0004-6361/202038960}

\bibitem[{{Soraisam} {et~al.}(2022){Soraisam}, {Matheson}, {Lee}, {Saha}, {Narayan}, {Wolf}, {Scott}, {Figuereo}, {Nu{\~n}ez}, {McKinnon}, {Guhathakurta}, {Brink}, {Filippenko}, \& {Smith}}]{Soraisam2022}
{Soraisam}, M., {Matheson}, T., {Lee}, C.-H., {et~al.} 2022, \apjl, 926, L11, \dodoi{10.3847/2041-8213/ac4e99}

\bibitem[{{Steele} {et~al.}(2004){Steele}, {Smith}, {Rees}, {Baker}, {Bates}, {Bode}, {Bowman}, {Carter}, {Etherton}, {Ford}, {Fraser}, {Gomboc}, {Lett}, {Mansfield}, {Marchant}, {Medrano-Cerda}, {Mottram}, {Raback}, {Scott}, {Tomlinson}, \& {Zamanov}}]{lt}
{Steele}, I.~A., {Smith}, R.~J., {Rees}, P.~C., {et~al.} 2004, in Society of Photo-Optical Instrumentation Engineers (SPIE) Conference Series, Vol. 5489, Ground-based Telescopes, ed. J.~{Oschmann}, Jacobus~M., 679--692, \dodoi{10.1117/12.551456}

\bibitem[{{Taddia} {et~al.}(2015){Taddia}, {Sollerman}, {Leloudas}, {Stritzinger}, {Valenti}, {Galbany}, {Kessler}, {Schneider}, \& {Wheeler}}]{Taddia2015A&A...574A..60T}
{Taddia}, F., {Sollerman}, J., {Leloudas}, G., {et~al.} 2015, \aap, 574, A60, \dodoi{10.1051/0004-6361/201423915}

\bibitem[{{Taddia} {et~al.}(2016){Taddia}, {Fremling}, {Sollerman}, {Corsi}, {Gal-Yam}, {Karamehmetoglu}, {Lunnan}, {Bue}, {Ergon}, {Kasliwal}, {Vreeswijk}, \& {Wozniak}}]{Taddia2016A&A...592A..89T}
{Taddia}, F., {Fremling}, C., {Sollerman}, J., {et~al.} 2016, \aap, 592, A89, \dodoi{10.1051/0004-6361/201628703}

\bibitem[{{Taddia} {et~al.}(2018){Taddia}, {Sollerman}, {Fremling}, {Karamehmetoglu}, {Quimby}, {Gal-Yam}, {Yaron}, {Kasliwal}, {Kulkarni}, {Nugent}, {Smadja}, \& {Tao}}]{Taddia2018}
{Taddia}, F., {Sollerman}, J., {Fremling}, C., {et~al.} 2018, \aap, 609, A106, \dodoi{10.1051/0004-6361/201629874}

\bibitem[{{Takei} \& {Tsuna}(2024)}]{TakeiTsuna2024}
{Takei}, Y., \& {Tsuna}, D. 2024, The Open Journal of Astrophysics, 7, 119, \dodoi{10.33232/001c.127610}

\bibitem[{{Tartaglia} {et~al.}(2021){Tartaglia}, {Sollerman}, {Barbarino}, {Taddia}, {Mason}, {Berton}, {Taggart}, {Bellm}, {De}, {Frederick}, {Fremling}, {Gal-Yam}, {Golkhou}, {Graham}, {Ho}, {Hung}, {Kaye}, {Kim}, {Laher}, {Masci}, {Perley}, {Porter}, {Reiley}, {Riddle}, {Rusholme}, {Soumagnac}, \& {Walters}}]{Tartaglia2021}
{Tartaglia}, L., {Sollerman}, J., {Barbarino}, C., {et~al.} 2021, \aap, 650, A174, \dodoi{10.1051/0004-6361/202039068}

\bibitem[{{Taubenberger} {et~al.}(2009){Taubenberger}, {Valenti}, {Benetti}, {Cappellaro}, {Della Valle}, {Elias-Rosa}, {Hachinger}, {Hillebrandt}, {Maeda}, {Mazzali}, {Pastorello}, {Patat}, {Sim}, \& {Turatto}}]{Taubenberger2009MNRAS}
{Taubenberger}, S., {Valenti}, S., {Benetti}, S., {et~al.} 2009, \mnras, 397, 677, \dodoi{10.1111/j.1365-2966.2009.15003.x}

\bibitem[{{Tonry} {et~al.}(2018){Tonry}, {Denneau}, {Heinze}, {Stalder}, {Smith}, {Smartt}, {Stubbs}, {Weiland}, \& {Rest}}]{atlas}
{Tonry}, J.~L., {Denneau}, L., {Heinze}, A.~N., {et~al.} 2018, \pasp, 130, 064505, \dodoi{10.1088/1538-3873/aabadf}

\bibitem[{van~der Walt {et~al.}(2011)van~der Walt, Colbert, \& Varoquaux}]{numpy}
van~der Walt, S., Colbert, S.~C., \& Varoquaux, G. 2011, Computing in Science \& Engineering, 13, 22, \dodoi{10.1109/MCSE.2011.37}

\bibitem[{van~der Walt {et~al.}(2019)van~der Walt, Crellin-Quick, \& Bloom}]{skyportal2019}
van~der Walt, S.~J., Crellin-Quick, A., \& Bloom, J.~S. 2019, Journal of Open Source Software, 4, \dodoi{10.21105/joss.01247}

\bibitem[{{Virtanen} {et~al.}(2020){Virtanen}, {Gommers}, {Oliphant}, {Haberland}, {Reddy}, {Cournapeau}, {Burovski}, {Peterson}, {Weckesser}, {Bright}, {van der Walt}, {Brett}, {Wilson}, {Millman}, {Mayorov}, {Nelson}, {Jones}, {Kern}, {Larson}, {Carey}, {Polat}, {Feng}, {Moore}, {VanderPlas}, {Laxalde}, {Perktold}, {Cimrman}, {Henriksen}, {Quintero}, {Harris}, {Archibald}, {Ribeiro}, {Pedregosa}, {van Mulbregt}, \& {SciPy 1. 0 Contributors}}]{scipy2020}
{Virtanen}, P., {Gommers}, R., {Oliphant}, T.~E., {et~al.} 2020, Nature Methods, 17, 261, \dodoi{10.1038/s41592-019-0686-2}

\bibitem[{{Woosley}(2017)}]{2017ApJ...836..244W}
{Woosley}, S.~E. 2017, \apj, 836, 244, \dodoi{10.3847/1538-4357/836/2/244}

\bibitem[{{Yang} \& {Sollerman}(2023)}]{YangHaffet}
{Yang}, S., \& {Sollerman}, J. 2023, \apjs, 269, 40, \dodoi{10.3847/1538-4365/acfcb4}

\bibitem[{{Yaron} \& {Gal-Yam}(2012)}]{wiserep}
{Yaron}, O., \& {Gal-Yam}, A. 2012, \pasp, 124, 668, \dodoi{10.1086/666656}

\bibitem[{{Zhu} {et~al.}(2024){Zhu}, {Liu}, {Yu}, {Mandel}, {Hirai}, {Zhang}, \& {Chen}}]{Zhu2024ApJ...970L..42Z}
{Zhu}, J.-P., {Liu}, L.-D., {Yu}, Y.-W., {et~al.} 2024, \apjl, 970, L42, \dodoi{10.3847/2041-8213/ad63a8}

\end{thebibliography}
\bibliographystyle{aasjournal}

\appendix 
\section{Photometry Data}\label{app:phot}

\begin{table}[H]
    \centering
    \footnotesize
    \caption{Log of optical photometry of SN 2021uvy and SN 2022hgk of $>5\sigma$ significance and corrected for MW extinction (full table available \href{10.5281/zenodo.15786071}{online}, DOI:10.5281/zenodo.15786071)}
    \label{tab:app:optical}
    \begin{tabular}{ccccc}
    \toprule
    \toprule
        IAU Name & MJD & Filter & Telescope & Brightness\\
         &   &    &    & (mag)     \\
    \midrule
SN~2021uvy & 59401.44 & $r$ & P48:ZTF & 20.85 $\pm$ 0.24 \\
... & & & \\
SN~2022hgk & 59672.32 & $g$ & P48:ZTF & 21.56 $\pm$ 0.22 \\
... & & & \\
\bottomrule
\end{tabular}
\end{table}

\begin{table}[H]
    \centering
    \footnotesize
    \caption{Log of UVOT observations of SN 2022hgk of $>3\sigma$ significance and corrected for MW extinction (full table available \href{10.5281/zenodo.15786071}{online}, DOI:10.5281/zenodo.15786071)}
    \label{tab:app:xrt}
    \begin{tabular}{ccc}
    \toprule
    \toprule
        MJD & Filter & Brightness\\
            &        & (mag)     \\
    \midrule

59720.72 & uvw2 & 19.625 $\pm$ 0.075 \\
... & & \\
\bottomrule
\end{tabular}
\end{table}

\section{Spectroscopy Data}\label{app:spec}

\begin{table}[h]
    \caption{Summary of optical spectra of SNe~2021uvy, 2022hgk, and 2020acct. We report phases (in rest-frame days) calculated with respect to both the first peak of the light curve and the estimated explosion epoch (inside parentheses).}
    \label{tab:app:speclist}
    \centering
    \begin{tabular}{ccccc||ccccc}
    \toprule
    \toprule
     IAU Name  &  MJD  & Phase & Telescope & Exposure & IAU Name  &  MJD  & Phase & Telescope & Exposure \\
      &  & (day) & /Instrument & (s) &  &  & (day) & /Instrument & (s) \\
        \midrule
SN~2020acct & 59195 & -1 (1)     & P60/SEDM   & 2700 & SN~2022hgk  & 59708 & 17 (33)   & LT/SPRAT   & 600   \\
            & 59253 & 55 (58)    & P200/DBSP  & 450  &             & 59709 & 18 (34)   & P60/SEDM   & 2700  \\
            & 59254 & 56 (59)    & P60/SEDM   & 2700 &             & 59710 & 19 (35)   & NOT/ALFOSC & 3600  \\
            & 59255 & 57 (60)    & NOT/ALFOSC & 1350 &             & 59713 & 22 (38)   & P60/SEDM   & 2700  \\
            & 59256 & 58 (61)    & P60/SEDM   & 2700 &             & 59718 & 27 (43)   & P60/SEDM   & 2700  \\
            & 59260 & 62 (65)    & NOT/ALFOSC & 900  &             & 59719 & 28 (44)   & P60/SEDM   & 2700  \\
            & 59260 & 62 (65)    & P60/SEDM   & 2700 &             & 59721 & 30 (46)   & NOT/ALFOSC & 1800  \\
            & 59263 & 64 (67)    & Keck1/LRIS & 1275 &             & 59722 & 31 (47)   & P200/DBSP  & 1200  \\
            & 59277 & 78 (81)    & NOT/ALFOSC & 1800 &             & 59730 & 39 (55)   & P60/SEDM   & 2700  \\
            & 59311 & 112 (114)  & Keck1/LRIS & 2312 &             & 59732 & 40 (56)   & NOT/ALFOSC & 2400  \\
            & 59350 & 149 (152)  & Keck1/LRIS & 2705 &             & 59734 & 43 (59)   & P60/SEDM   & 2700  \\
            &       &            &            &      &             & 59738 & 46 (62)   & P200/DBSP  & 1500  \\
SN~2021pkd  & 59386 & -7 (12)    & P60/SEDM   & 2700 &             & 59739 & 47 (63)   & P200/DBSP  & 900   \\ 
            & 59389 & -4 (15)    & P60/SEDM   & 2700 &             & 59788 & 95 (111)  & Keck1/LRIS & 900   \\
            & 59391 & -3 (17)    & P60/SEDM   & 2700 &             &       &           &            &       \\
            & 59401 &  7 (26)    & Keck1/LRIS & 300  & SN~2023plg  & 60242 & 70 (70)   & P60/SEDM   & 2700  \\
            &       &            &            &      &             & 60246 & 74 (74)   & LT/SPRAT   & 750   \\
SN~2021uvy  & 59439 & $-15$ (38) & NTT/EFOSC2 & 900  &             & 60246 & 74 (74)   & P60/SEDM   & 2160  \\
            & 59442 & $-12$ (40) & P60/SEDM   & 2700 &             & 60249 & 77 (77)   & P60/SEDM   & 2160  \\
            & 59454 & $-1$ (51)  & P60/SEDM   & 2700 &             & 60254 & 82 (82)   & P60/SEDM   & 2160  \\
            & 59455 & 0 (52)     & LT/SPRAT   & 750  &             & 60256 & 84 (84)   & Keck1/LRIS & 300   \\
            & 59458 & 2 (55)     & P200/DBSP  & 600  &             & 60259 & 87 (87)   & P60/SEDM   & 2160  \\
            & 59467 & 10 (62)    & Keck1/LRIS & 600  &             & 60269 & 97 (97)   & P60/SEDM   & 2700  \\
            & 59467 & 11 (63)    & P60/SEDM   & 2700 &             & 60274 & 102 (102) & P60/SEDM   & 2700  \\
            & 59470 & 13 (66)    & P200/DBSP  & 900  &             & 60275 & 102 (102) & NOT/ALFOSC & 2400  \\
            & 59491 & 33 (85)    & Keck1/LRIS & 600  &             & 60280 & 107 (107) & P60/SEDM   & 2760  \\
            & 59498 & 39 (91)    & P60/SEDM   & 2700 &             & 60281 & 108 (108) & P60/SEDM   & 2760  \\
            & 59502 & 43 (95)    & P60/SEDM   & 2700 &             & 60281 & 108 (108) & P200/DBSP  & 1200  \\
            & 59509 & 49 (101)   & P60/SEDM   & 2700 &             & 60282 & 109 (109) & P60/SEDM   & 3624  \\
            & 59517 & 57 (109)   & P60/SEDM   & 2700 &             & 60283 & 110 (110) & P60/SEDM   & 396   \\
            & 59524 & 63 (115)   & P200/DBSP  & 900  &             & 60285 & 112 (112) & P60/SEDM   & 2700  \\
            & 59536 & 74 (126)   & P60/SEDM   & 2700 &             & 60285 & 112 (112) & P60/SEDM   & 3840  \\
            & 59547 & 84 (136)   & P60/SEDM   & 2700 &             & 60286 & 113 (113) & NOT/ALFOSC & 2400  \\
            & 59561 & 97 (149)   & P60/SEDM   & 2700 &             & 60288 & 115 (115) & P60/SEDM   & 2700  \\
            & 59585 & 118 (171)  & P60/SEDM   & 2700 &             & 60296 & 123 (123) & P60/SEDM   & 2700  \\
            & 59587 & 120 (172)  & NOT/ALFOSC & 2700 &             & 60299 & 126 (126) & NOT/ALFOSC & 1200  \\
            & 59600 & 132 (185)  & P60/SEDM   & 2700 &             & 60321 & 147 (147) & NOT/ALFOSC & 2400  \\
            & 59615 & 146 (198)  & Keck1/LRIS & 300  &             &       &           &            &       \\
            & 59815 & 328 (380)  & Keck1/LRIS & 1800 &             &       &           &            &       \\ 
            & 59875 & 384 (436)  & Keck1/LRIS & 2700 &             &       &           &            &       \\ 
        \bottomrule
    \end{tabular}
\end{table}
\end{document}